\documentclass[]{aastex63}

\usepackage[version=4]{mhchem}
\usepackage{amsmath,amssymb}
\usepackage{braket} 
\usepackage{etoolbox}
\usepackage{lineno}
\received{}
\revised{}
\accepted{}
\submitjournal{ApJ}

\shorttitle{\ce{H2} nuclear spin conversion II}
\shortauthors{Furuya et al.}
\newcommand{\edes}{{E_{\rm des}}}

\newcommand{\ehop}{{E_{\rm hop}}}

\newcommand{\phh}{{\rm p\mathchar`- H_{2}}}
\newcommand{\ohh}{{\rm o\mathchar`- H_{2}}}

\newcommand{\op}[1]{{\rm OPR(\ce{#1})}}
\newcommand{\opst}{{\rm OPR_{st}}}

\newcommand{\deops}{\Delta E_{\rm op,\,s}}
\newcommand{\deopg}{\Delta E_{\rm op,\,g}}

\begin{document}

\title{\ce{H2} Ortho–Para Spin Conversion on Inhomogeneous Grain Surfaces. II. impact of the rotational energy difference between adsorbed ortho-\ce{H2} and para-\ce{H2} and implication to deuterium fractionation chemistry}

\correspondingauthor{Kenji Furuya}
\email{kenji.furuya@riken.jp}

\author[0000-0002-2026-8157]{Kenji Furuya}
\affiliation{RIKEN Pioneering Research Institute, 2-1 Hirosawa, Wako-shi, Saitama 351-0198, Japan}
\author[0000-0003-3453-6009]{Toshiki Sugimoto}
\affiliation{Department of Materials Molecular Science, Institute for Molecular Science, Myodaiji, Okazaki, Aichi 444-8585, Japan}
\author[0000-0002-2026-8157]{Kazunari Iwasaki}
\affiliation{Center for Computational Astrophysics/Division of Science, National Astronomical Observatory of Japan, Mitaka, Tokyo 181-8588, Japan}
\affiliation{Astronomical Science Program, The Graduate University for Advanced Studies (SOKENDAI), 2-21-1 Osawa, Mitaka, Tokyo 181-8588, Japan}
\author[0000-0001-9669-1288]{Masashi Tsuge}
\affiliation{Institute of Low Temperature Science, Hokkaido University, Sapporo, Hokkaido 060-0819, Japan}
\author[0000-0001-8408-2872]{Naoki Watanabe}
\affiliation{Institute of Low Temperature Science, Hokkaido University, Sapporo, Hokkaido 060-0819, Japan}



\begin{abstract}
%We show that nitrogen can become the third most abundant element in the gas phase after hydrogen and helium in the warm molecular layers of protoplanetary disks.
We investigate how the \ce{H2} ortho-to-para ratio (OPR) and dueterium fractionation in star-forming regions are affected by nuclear spin conversion (NSC) on dust grains.
Particular focus is placed on the rotational energy difference between ortho-\ce{H2} (o-\ce{H2}) and para-\ce{H2} (p-\ce{H2}) on grain surfaces.
While the ground state of o-\ce{H2} has a higher rotational energy than that of p-\ce{H2} by 170.5 K in the gas phase, this energy difference is expected to become smaller on solid surfaces, where interactions between the surface and adsorbed \ce{H2} molecules affect their rotational motion.
A previous study by \citet{furuya19} developed a rigorous formulation of the rate for the temporal variation of the \ce{H2} OPR via the NSC on grains, assuming that adsorbed o-\ce{H2} has higher rotational energy than adsorbed p-\ce{H2} by 170.5 K, as in the gas phase.
In this work, we relax the assumption and re-evaluate the rate, varying the rotational energy difference between their ground states.
The re-evaluated rate is incorporated into a gas-ice astrochemical model to study the evolution of the \ce{H2} OPR and the deuterium fractionation in prestellar cores and the outer, cold regions of protostellar envelopes.
The inclusion of the NSC on grains reduces the timescale of the \ce{H2} OPR evolution and thus the deuterium fractionation, at densities of $\gtrsim$10$^4$ cm$^{-3}$ and  temperatures of $\lesssim$14-16 K (depending on the rotational energy difference), when the ionization rate of \ce{H2} is 10$^{-17}$ s$^{-1}$.
\end{abstract}

\keywords{astrochemistry --- ISM: molecules}




\section{Introduction}
%$\tau_{\rm VCF} = \chi (H/D_z)^2$
The deuterium fractionation ratio of molecules in star- and planet-forming regions is a useful tool to investigate where and how the molecules were formed, and to investigate the possible chemical link between the interstellar molecules and the primitive materials in the solar system \citep[e.g.,][]{ceccarelli14,nomura23}.
Elemental chemical processes relevant to the deuterium fractionation in the gas phase and on the solid surface have been extensively studied by laboratory experiments and quantum chemical calculations \citep[e.g.,][]{watson76,nagaoka05,hidaka09,ratajczak09,roueff13,cooper19,hillenbrand19,Jimenez-Redondo24}.
It is well established that the starting point of the deuterium fractionation is the isotope exchange reaction,
\begin{linenomath*}
\begin{align}
\ce{H3+} + \ce{HD} \rightarrow \ce{H2D+} + \ce{H2} + \Delta E. \label{eq:ex1}
\end{align}
\end{linenomath*}
At a typical temperature of molecular clouds, $\sim$10 K, the backward reaction is inefficient due to the endothermicity (but see below) and thus \ce{H2D+} becomes abundant with time
with respect to \ce{H3+}.
%\ce{H2D+} can further react with HD to form \ce{D2H+} and \ce{D3+} \citep{roberts03}:
%\begin{align}
%&\ce{H2D+} + \ce{HD} \rightarrow \ce{D2H+} + \ce{H2} + \Delta E_2, \label{eq:ex2} \\
%&\ce{D2H+} + \ce{HD} \rightarrow \ce{D3+} + \ce{H2} + \Delta E_3. \label{eq:ex3}
%\end{align}
Since \ce{H3+} plays a central role in the interstellar chemistry, the deuterium enrichment in \ce{H3+} is transferred to other gaseous and icy molecules \citep[e.g.,][]{tielens83,brown89}.
As a result, all molecules except for \ce{H2} are expected to have a higher D/H ratio than the elemental deuterium-to-hydrogen ratio of $1.5\times10^{-5}$ \citep{linsky03}.
Indeed, observations have found orders of magnitude enhancement in the D/H ratio of various molecules not only in molecular clouds but also in protostellar sources and protoplanetary disks \citep[see][for recent reviews]{ceccarelli14,nomura23}.

\begin{figure*}[ht!]
\plotone{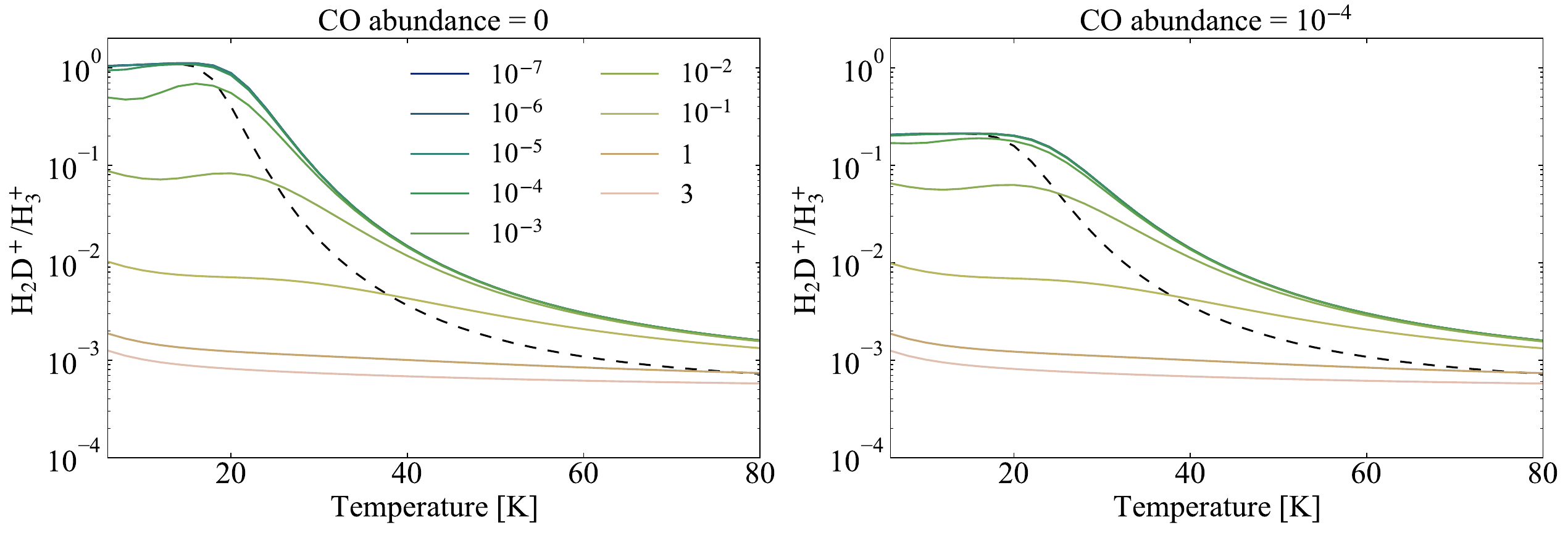}
\caption{The steady-state value of the \ce{H2D+}/\ce{H3+} abundance ratio as functions of the gas temperature, varying in the \ce{H2} OPR in the range between 10$^{-7}$ and 3 (solid lines). Dashed lines show the \ce{H2D+}/\ce{H3+} ratio assuming a thermalized value for the \ce{H2} OPR. i.e., $9\exp(-170.5/T_g)$. The gas-phase CO abundance is assumed to be zero in the left panel, while it is assumed to be 10$^{-4}$ in the right panel.
These two cases correspond to the minimum and maximum limits of CO abundance, and in reality, the CO abundance would fall between these extremes.
}
\label{fig:analytical}
\end{figure*}

\ce{H2} and \ce{H3+} have two nuclear spin configurations, ortho and para;
the nuclear-spin wave functions of ortho and para states are symmetric and antisymmetric, respectively, with
respect to the permutations of the nuclei.
The endothermicity of the isotope exchange reaction (\ref{eq:ex1}) in the reverse direction ($\Delta E$) depends on the nuclear spin state of \ce{H2} and that of \ce{H3+} isotopologues.
Thus, their ortho-to-para ratios (OPRs) are the critical parameters for the overall deuterium fractionation chemistry \citep[e.g.,][]{pagani92,flower06}.
The OPRs of \ce{H3+} and \ce{H2D+} are mostly determined by the proton exchange with \ce{H2}, and given as functions of the \ce{H2} OPR and the gas temperature ($T_g$) \citep[][]{gerlich02}.
Thus, the \ce{H2} OPR is of primary importance.
Figure \ref{fig:analytical} shows the steady-state \ce{H2D+}/\ce{H3+} abundance ratio
%, \ce{D2H+}/\ce{H2D+}, and \ce{D3+}/\ce{D2H+} ratios 
as a function of $T_g$, varying in the \ce{H2} OPR.
When the \ce{H2} OPR is greater than $\sim$10$^{-3}$, \ce{H2} rather than CO is the dominant destroyer of \ce{H2D+}, regardless of the CO abundance, and the degrees of the deuterium fractionation depend on the \ce{H2} OPR \citep[see also][]{furuya15}. 
%Then a detailed understanding of the \ce{H2} OPR is required for accurate modeling of the deuterium fractionation chemistry.

In the ISM, the \ce{H2} OPR is determined by the competition among the three processes: the \ce{H2} formation via recombination of two H atoms on grain surfaces \citep{watanabe10,tsuge19}, the proton exchange reactions between \ce{H2} and \ce{H+} (or \ce{H3+}) in the gas phase \citep{gerlich90,honvault11}, and the nuclear spin conversion (NSC) on solid surfaces \citep{ueta16,tsuge21a,tsuge21b}.
Astrochemical models with deuterium fractionation chemistry often consider the \ce{H2} formation and the NSC of \ce{H2} via gas-phase reactions, but commonly neglect the NSC on grain surfaces \citep[e.g.,][]{flower06,sipila13,taquet14,furuya15,hilyblant18}.
One exception is \citet{bovino17}, but they considered the NSC of ortho-\ce{H2} ($\ohh$) to para-\ce{H2} ($\phh$) only, neglecting the conversion in the reverse direction and assuming a single value for \ce{H2} binding energy.
Later \citet{furuya19} pointed out the importance of the \ce{H2} binding energy distribution. 
They formulated the rates for the temporal evolution of the gas-phase \ce{H2} OPR via NSC on water ice under the ISM conditions, based on the laboratory-measured NSC conversion timescale and the \ce{H2} binding energy distribution.
In their formulation, the rotational energy difference between the ground state of o-\ce{H2} and p-\ce{H2} was assumed to be the same in the gas phase ($\deopg=170.5$ K) and on solid surfaces ($\deops$).
However, on solid surfaces, the rotational motion of \ce{H2} molecules is suppressed due to the interaction with the surface and $\deops$ would be smaller than $\deopg$ (see Section \ref{sec:energy_difference} and the Appendix for details).
In this work, we extend the approach by \citet{furuya19} to re-evaluate the rates for the temporal evolution of the \ce{H2} OPR via NSC on surfaces in the case when $\deops \leq \deopg$.
Moreover, by adding the NSC on surfaces to a gas-ice astrochemical model with deuterium fractionation chemistry \citep{furuya15}, we address how the NSC on surfaces affects the deuterium fractionation chemistry in prestellar cores and in protostellar envelopes, treating $\deops$ as a free parameter. 
%As already mentioned above, the actual value of $E_{\rm op,\,s}$ remains unknown, so it is treated as a free parameter in this work.
%For implementing the NSC on solid surfaces to astrochemical models with deuterium chemistry, there are two major problems to be solved.
%Firstly, as already mentioned above, the actual value of $E_{\rm op,\,s}$ remains unknown, and it is unclear how the uncertainty affects the \ce{H2} OPR evolution.
%Secondly, the binding energy distribution of \ce{H2} does matter for the NSC on grains, while astrochemical models with a large chemical network commonly neglect the binding energy distribution.

%In this work, we show that the value of $E_{\rm op,\,s}$ affects the \ce{H2} OPR on grain surfaces, but it does not affect the evolution of \ce{H2} OPR in the bulk gas (Section \ref{sec:sec2}).
%We also propose a new method for treating \ce{H2} on grains based on quasi steady-state assumption (Appendix).
The rest of this paper is organized as follows.
In Section \ref{sec:sec2}, we discuss how the $\deops$ value affects the evolution of \ce{H2} OPR in the bulk gas under the ISM conditions, and re-evaluate the rates for the temporal evolution of the \ce{H2} OPR via NSC on surfaces under the ISM conditions.
The timescale of the \ce{H2} OPR evolution and its steady state value in the ISM are discussed in Section \ref{sec:steady_state}.
In Section \ref{sec:model}, we discuss the impact of the NSC on grain surfaces on the deuterium fractionation in a collapsing prestellar core.
In Section \ref{sec:discuss}, we discuss the impact of the NSC on grain surfaces on the \ce{H2} OPR and \ce{H2D+} OPR in the outer regions of the protostellar envelope.
Our main findings are summarized in Section \ref{sec:summary}.

%The rest of this paper is organized as follows.
%In Section \ref{sec:steady_state}, we discuss how uncertainty in $\deops$ affects the steady-state value of \ce{H2} OPR in different physical conditions.
%In Section \ref{sec:model}, we describe our numerical model, which considers the time evolution of the deuterium %fractionation and the \ce{H2} OPR in a self-consistent way, and discuss numerical results in Section \ref{sec:result}.

\section{NSC of \ce{H2} on grain surfaces} \label{sec:sec2}
%, $\gamma_{\rm s} = 9 \exp(-\Delta E_{\rm op,\,s}/T_{\rm s})$,
%where $\Delta E_{\rm op,\,s}$ is $E_{J=1}-E_{J=0}$ on the solid surface, and $T_{\rm s}$ is the surface temperature.
%On solid surfaces, the rotational motion of \ce{H2} molecules is suppressed due to the interaction with the surface and $\deops$ would be smaller than $\deopg$ (see Section \ref{sec:energy_difference} for details).
%Thus $\gamma_{\rm s}$ can be larger than $\gamma_{\rm g}$, although the exact value of $\Delta E^{\rm s}_{\rm op}$ remains unclear.

%The gas-phase proton exchange reactions tend to change the \ce{H2} OPR closer to $\gamma_{\rm g}$ = $9 \exp(-\deopg/T_g)$, where $\deopg$ is $E_{J=1}-E_{J=0}$ in the gas phase, where \ce{H2} molecules can rotate freely.
%Similarly, the NSC on solid surfaces would change the \ce{H2} OPR closer to the thermalized value, $\gamma_{\rm s} = 9 \exp(-\Delta E_{\rm op,\,s}/T_{\rm s})$,
%where $\Delta E_{\rm op,\,s}$ is $E_{J=1}-E_{J=0}$ on the solid surface, and $T_{\rm s}$ is the surface temperature.
%On solid surfaces, the rotational motion of \ce{H2} molecules is suppressed due to the interaction with the surface and $\deops$ would be smaller than $\deopg$ (see Section \ref{sec:energy_difference} for details).
%Thus $\gamma_{\rm s}$ can be larger than $\gamma_{\rm g}$, although the exact value of $\Delta E^{\rm s}_{\rm op}$ remains unclear.

\subsection{Rotational energy difference between ortho and para hydrogen adsorbed on grains} \label{sec:energy_difference}
The \ce{H2} OPR in thermal equilibrium is given by 
\begin{align}
\frac{ \displaystyle \sum_{J = \rm{odd}} \displaystyle \sum_{m=-J}^{J} d_{I=1} \exp(-E_{J,\,m}/T)}{\displaystyle \sum_{J = \rm even} \displaystyle \sum_{m=-J}^{J} d_{I=0} \exp(-E_{J,\,m}/T)}, \label{eq:h2opr_def1}
\end{align}
where $d_{I=1}$ = 3 and $d_{I=0}$ = 1 are the nuclear-spin degeneracies and $T$ is the temperature.
$E_{J,\,m}$ is the rotational energy of $(J, m)$ state in Kelvin, where $J$ is the rotational quantum number, and $m$ is the quantum number associated with the $z$-component of the angular momentum.
For gas-phase \ce{H2}, which can rotate freely, the rotational energy does not depend on $m$, then Eq. \ref{eq:h2opr_def1} can be rewritten as
\begin{align}
\frac{ \displaystyle \sum_{J = \rm{odd}} d_{I=1} (2J+1) \exp(-E_{J}/T)}{\displaystyle \sum_{J = \rm even} \displaystyle d_{I=0} (2J+1) \exp(-E_{J}/T)}, \label{eq:h2opr_def2}
\end{align}
where $2J+1$ is the rotational degeneracy \citep[e.g.,][]{fukutani13}.
In the case of three-dimensional free rotation, the rotational energy is given as $E_J = BJ(J+1)$, where $B$ is the rotational constant ($\sim$7.4 meV or $\sim$85 K for \ce{H2}).
Then the rotational energy difference between the lowest rotational level of p-\ce{H2} ($J=0$) and o-\ce{H2} ($J=1$) is $\deopg = 2 B \sim$170 K.
At the low temperature limit ($\lesssim$80 K), where only the ground states of ortho-\ce{H2} ($\ohh$) and para-\ce{H2} ($\phh$) are relevant and the population of higher rotational states ($J \geq 2$) is negligible, the thermalized value for freely rotating \ce{H2} is approximated by
\begin{align}
\gamma_g = 9 \exp(-\deopg/T_g),
\end{align}
where $T_g$ is the gas temperature.
The gas-phase proton exchange reactions tend to change the \ce{H2} OPR closer to $\gamma_{g}$ \citep{gerlich90,le_Bourlot91}.
Similarly, the NSC on solid surfaces would change the \ce{H2} OPR closer to the thermalized value on the surfaces, which is discussed below.
In \citet{furuya19}, the thermalized value on the surfaces was assumed to be $9\exp(-\deopg/T_s)$, where $T_s$ is the surface temperature.
Throughout this work, we consider only the ground states of o-\ce{H2} ($J=1$) and p-\ce{H2} ($J=0$), unless otherwise stated.

In proximity to surfaces, the surface-molecule interaction influences the rotational motion, and the rotational symmetry is broken in general. Whereas heavy molecules stop rotating on surfaces with the molecular axis fixed to the surface, the lightest molecule like \ce{H2} retains its rotational motion even under the anisotropic surface potential due to the quantum effect \citep{fukutani13}. 
Under the anisotropic surface potential, the degeneracy of the $J = 1$ state is lifted and $(J, m)$ = (1, 0) and (1, $\pm1$) states are shifted, while the $J = 0$ state shows no shift (Figure \ref{fig:erot} and see Appendix for more details).
Which of the (1, 0) or (1, $\pm1$) states becomes the lowest in energy for o-\ce{H2} depends on the property of anisotropic surface potential.
Throughout this work, we assume that (1, $\pm1$) states, which correspond to the case of a two-dimensional hindered rotor, are the ground states of adsorbed o-\ce{H2}; i.e., the $V_2>0$ case given in Fig. \ref{fig:erot}.
In addition, we neglect the (1, 0) state for simplicity, because at the low temperatures ($\lesssim$20 K), most o-\ce{H2} is expected to occupy the ground (1, $\pm1$) state.

Under these assumptions, we express the thermalized value of OPR for adsorbed \ce{H2} as 
\begin{align}
\gamma_s = 6\exp(-\deops/T_s),
\end{align}
where $\deops$ corresponds to the energy difference between $(J,m)=(0,0)$ and $(1,\pm1)$ states on surfaces.
If we choose $(J, m) = (1,0)$ as the ground state of adsorbed o-\ce{H2}, the pre-exponential factor becomes three.
The value of $\deops$ is much more important for $\gamma_s$ than that of the pre-exponential factor. 

Although the actual value of $\deops$ is unknown, the allowed values of $\deops$ can be constrained to some extent.
Under the surface potential with extremely large rotational anisotropy, the rotational motion tends to be restricted to the surface-parallel direction.
The rotational energy for two-dimensional hindered rotation is given as $BJ^2$.
Then the rotational energy difference between the ground state of p-\ce{H2} and o-\ce{H2} becomes $\Delta E_{\rm op,\,2Drot} = B \sim$ 85 K, which is about a half of $\deopg$.
In reality, where the surface potential does not have extremely large rotational anisotropy, $\deops$ would be somewhere between $\deopg$ and $\Delta E_{\rm op,\,2Drot}$ ($\Delta E_{\rm op,\,2Drot} \le \Delta E_{\rm op,\,s} \le \deopg$; see Appendix for more detail).
For example, assuming $\Delta E_{\rm op,\,s} = \Delta E_{\rm op,\,2Drot}$ and $T_{\rm g} = T_{s} = 10$ K, 
$\gamma_{s}$ is $1\times10^{-3}$, which is much larger than $\gamma_{g}$ at 10 K ($4\times10^{-7}$).

\begin{figure*}[ht!]
\plotone{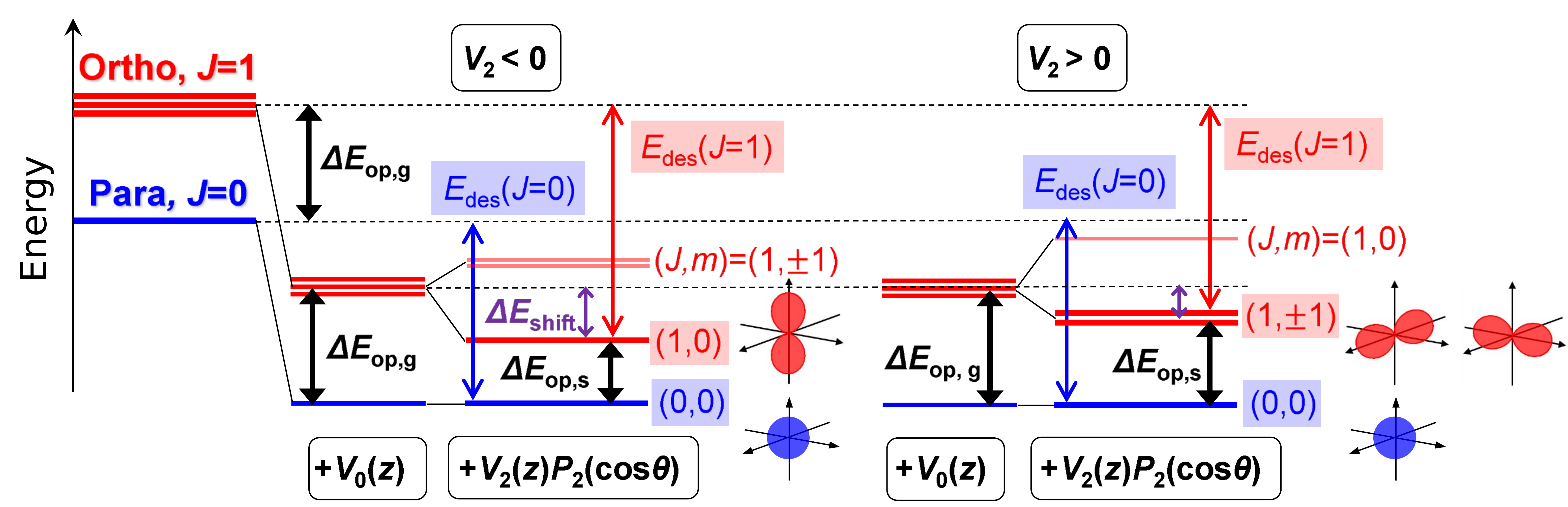}
\caption{Energy level diagram for p-\ce{H2} ($J=0$) and o-\ce{H2} ($J=1$) rotational states in the gas phase and on solid surfaces when the surface potential is given by $V(z,\theta)=V_0(z)+V_2P_2(\cos \theta)$, where $P_2(\cos \theta)$ is the second order Legendre polynomial, and $V_0(z)$ and $V_2$ are the isotropic surface potential and the degree of anisotropy, respectively. The number of lines in each level means the rotational degeneracy. Schematic illustrations show the probability distribution of the molecular axis for the respective states, $|Y_{0,0}|^2$, $|Y_{1,0}|^2$, $|Y_{1,1}|^2$, and $|Y_{1,-1}|^2$, where $Y_{J,m}$ is the spherical harmonics. See Appendix for more details.}
\label{fig:erot}
\end{figure*}

\subsection{Binding energy difference between ortho and para hydrogen}
As most \ce{H2} is present in the gas phase in the ISM rather than on grains, the exchange of gaseous and icy \ce{H2} is the rate-limiting step for the temporal evolution of the gas-phase \ce{H2} OPR via the NSC on grain surfaces \citep{furuya19}.
In addition, the efficiency of nuclear spin state conversion for an adsorbed \ce{H2} molecule prior to desorption depends on its residence time on the surface.
Then, alongside $\deops$, another important parameter would be the binding energy difference between o-\ce{H2} and p-\ce{H2}.
As o-\ce{H2} is sensitive to the anisotropic part of the surface potential, while p-\ce{H2} is not, o-\ce{H2} has a higher binding energy than p-\ce{H2} by $\deopg - \deops$ \citep[][see Fig. \ref{fig:erot}]{fukutani13}. 
Then, the \ce{H2} OPR on the surface may not be directly reflected in the gas phase after thermal desorption, because of the smaller desorption rate of o-\ce{H2} than p-\ce{H2}.
Actually, in the case when a single binding energy characterizes a surface,
the two effects ($\deops \leq \deopg$ and higher binding energy of o-\ce{H2} than p-\ce{H2}) are canceled out;
$\gamma_s \times \exp(-(\deopg - \deops)/T_s) \approx \gamma_g$.
In reality, however, surfaces contain various binding sites with different potential energy depths, and \ce{H2} can hop between different sites by thermal hopping.
Then, it is not obvious how the value of $\Delta E_{\rm op,\,s}$ affects \ce{H2} OPR in the gas phase after thermal desorption under the ISM conditions.
%singleの場合は打ち消しあう。multiple binding siteの場合はどうか。

\subsection{Numerical model} \label{sec:detail_model}
Here we explore how the value of $\deops$ affects the temporal evolution of the \ce{H2} OPR under the ISM conditions.
In the case where $\deops=\deopg$, the conversion rate from o-\ce{H2} to p-\ce{H2}
through the adsorption of o-\ce{H2}, subsequent NSC on grains, and the thermal desorption of p-\ce{H2} ($R_{\rm op, \,s}$ in a unit of cm$^{-3}$ s$^{-1}$) and the rate of the reverse process, $R_{\rm po, \,s}$, can be expressed as
\begin{linenomath*}
\begin{align}
R_{\rm op, \,s} &=  \eta_{\rm op}S_{\rm H_2}(1-\Theta)R_{\rm col}(\ohh), \label{eq:op_surf1} \\
R_{\rm po, \,s} &= \eta_{\rm po}S_{\rm H_2}(1-\Theta)R_{\rm col}(\phh), \label{eq:op_surf2}
%\frac{\eta_{\rm op}}{\eta_{\rm po}} &= 6\exp\left(-\frac{\deopg}{T_{\rm s}}\right), %\label{eq:op_surf3}
\end{align}
\end{linenomath*}
where $\eta_{\rm op}$ is the yield of gaseous $\phh$ per $\ohh$ adsorption,
$\eta_{\rm po}$ is the yield of gaseous $\ohh$ per $\phh$ adsorption,
$S_{\ce{H2}}$ is the sticking probability of \ce{H2} to the surface, $\Theta$ is the surface coverage of \ce{H2},
and $R_{\rm col}(\ohh)$ and $R_{\rm col}(\phh)$ are the collisional rates of $\ohh$ and $\phh$, respectively, to grains \citep{furuya19}.
The factor $1 - \Theta$ is considered because only one \ce{H2} molecule is assumed to be adsorbed per binding site.
$\eta_{\rm op}$ (and $\eta_{\rm po}$) describe the competition between the NSC of \ce{H2} on and the desorption of \ce{H2} from surfaces \citep{fukutani13,furuya19}.
%Equations \ref{eq:op_surf1} and \ref{eq:op_surf2} implicitly assume the adsorption–desorption equilibrium of \ce{H2}, which is well justified for molecular clouds or higher density regions (timescale to reach the equilibrium is $\sim$1 yr for 10$^{4}$ cm$^{-3}$) \citep{furuya19}.
The underlying assumption behind the formulation by \citet{furuya19} is that \ce{H2} is in the adsorption-desorption equilibrium.
As the adsorption and desorption of \ce{H2} reach the equilibrium in a very short timescale ($\sim$1 yr ($10^4$ cm$^{-3}$/$n(\ce{H2})$), the assumption should be valid in the dense ISM.
In addition to Eqs. \ref{eq:op_surf1} and \ref{eq:op_surf2}, the following relation was used in \citet{furuya19}:
\begin{linenomath*}
\begin{align}
\frac{\eta_{\rm op}}{\eta_{\rm po}} &= 9\exp\left(-\frac{\deopg}{T_{\rm s}}\right) = \gamma_{\rm g}. \label{eq:op_surf3}
\end{align}
\end{linenomath*}
Note that the ratio of $\eta_{\rm op}$ to $\eta_{\rm po}$ represents the equilibrium \ce{H2} OPR established solely by the NSC on grains.
Eqs. \ref{eq:op_surf1}-\ref{eq:op_surf3} can be incorporated in any astrochemical models based on the rate-equation approach.
One may think that when $\deops \neq \deopg$, one should simply replace $\deopg$ in Eq. \ref{eq:op_surf3} with $\deops$.
That would be true when a single binding energy for o-\ce{H2} and p-\ce{H2} is considered, and the binding energy of o-\ce{H2} and p-\ce{H2} is the same.
However, as mentioned above, it is not obvious when the binding energy has a distribution and o-\ce{H2} is more strongly bound on the surface.
This section aims to determine whether the formulation by \citet{furuya19} requires any modifications in the case when $\deops \neq \deopg$ 
and to re-evaluate $\eta_{\rm op}$ and $\eta_{\rm po}$.

\subsubsection{Model setup}
Following \citet{furuya19}, we numerically solve a set of ordinary differential equations for o-\ce{H2} and p-\ce{H2} that include the adsorption of gas-phase \ce{H2}, thermal desorption, thermal hopping, and NSC of adsorbed \ce{H2}, considering various binding sites with different potential energy depths on surfaces:
\begin{linenomath*}
\begin{align}
%\frac{d\num{\alpha \mathchar`- \ce{H2}}}{dt} &= - (1-\Theta(t))SR_{\rm col}(\alpha \mathchar`- \ce{H2}) + n_{\rm gr}N_{\rm site}\int k_{\rm thdes}(\edess)\theta_{\alpha}(\edess, t)g(\edess)d\edess, \label{eq:gas_h2} \\
%\frac{d\theta_{\alpha}(\edes, t)}{dt} &= \frac{1}{4}[1-\theta(\edes, t)]Sv_{\rm th} n_{\rm site}^{-1}\num{\alpha \mathchar`- \ce{H2}} - k_{\rm thdes}(\edes)\theta_{\alpha}(\edes, t)  \label{eq:cov_h2} \\ 
%&- \int k_{\rm hop}(\edes \rightarrow \edess) \theta_{\alpha}(\edes, t) [1-\theta(\edess, t)]g(\edess)d\edess \nonumber \\
%&+  [1-\theta(\edes, t)]\int k_{\rm hop}(\edess \rightarrow \edes) \theta_{\alpha}(\edess, t)g(\edess)d\edess \nonumber \\
%&+ k^{surf}_{\beta\alpha}\theta_{\beta}(\edes, t) - k^{surf}_{\alpha\beta}\theta_{\alpha}(\edes, t), \nonumber
\frac{d n(\alpha \mathchar`- \ce{H2})}{dt} &= - (1-\Theta(t))S_{\rm H_2}R_{\rm col}(\alpha \mathchar`- \ce{H2}) + n_{\rm gr}N_{\rm site}\Sigma_j [k_{\rm thdes}(j)\theta_{\alpha}(j, t)g(j)], \label{eq:gas_h2} \\
\frac{d\theta_{\alpha}(i, t)}{dt} &= [1-\theta(i, t)]S_{\rm H_2} \frac{\sigma_{\rm gr}}{N_{\rm site}} v_{\rm th} n(\alpha \mathchar`- \ce{H2}) - \Sigma_j k_{\rm thdes}(j)\theta_{\alpha}(j, t)  \label{eq:cov_h2} \\ 
&- \Sigma_j k_{\rm hop}(i \rightarrow j) \theta_{\alpha}(i, t) [1-\theta(j, t)]g(j) \nonumber \\
&+  [1-\theta(i, t)]\Sigma_j [k_{\rm hop}(j \rightarrow i) \theta_{\alpha}(j, t)g(j)] \nonumber \\
&+ k^{\rm surf}_{\beta\alpha}\theta_{\beta}(i, t) - k^{\rm surf}_{\alpha\beta}\theta_{\alpha}(i, t), \nonumber
\end{align}
\end{linenomath*}
where $\alpha$ and $\beta$ indicate either ortho ($o$) or para ($p$), and $n({\rm o} \mathchar`- \ce{H2})$ and $n({\rm p} \mathchar`- \ce{H2})$ indicate the number density of o-\ce{H2} and p-\ce{H2} in the gas-phase, respectively.
$n_{\rm gr}$ is the number density of dust grains in a unit gas volume, $N_{\rm site}$ is the number of binding sites on a grain surface, and $k_{\rm thdes}$ is the rate constant of the thermal desorption.
We denote the fraction of binding site $i$ on the surface as $g(i)$, which satisfies
\begin{linenomath*}
\begin{equation}
\Sigma_i g(i) = 1.
\end{equation}
\end{linenomath*}
We denote the fraction of binding site $i$ that are occupied by $\ohh$ ($\phh$) as $\theta_o(i)$ ($\theta_p(i)$), and $\theta(i)$ is defined by $\theta(i) = \theta_o(i) + \theta_p(i)$.
The surface coverage of \ce{H2} at a given time $t$, $\Theta(t)$, is defined as
\begin{linenomath*}
\begin{equation}
\Theta(t) = \Sigma_i \theta(i, t)g(i). \label{eq:Thetat}
\end{equation}
\end{linenomath*}
Similarly $\Theta_{\alpha}(t)$, where $\alpha$ is $o$ or $p$, is defined as
\begin{linenomath*}
\begin{equation}
\Theta_{\alpha}(t) = \Sigma_i \theta_{\alpha}(i, t)g(i), 
\end{equation}
\end{linenomath*}
and thus $\Theta(t) = \Theta_{o}(t) + \Theta_{p}(t)$.
%Eqs. (\ref{eq:gas_h2}) and (\ref{eq:cov_h2}) are similar to Eqs. (1) and (2) in \citet{furuya19} except that our equations distinguish between $\ohh$ and $\phh$.
%The collision rates to dust grains and desorption rates from the whole surface of dust grains of  $\ohh$ and $\phh$ are given by
%\begin{align}
%R_{\rm col}(\alpha \mathchar`- \ce{H2}) &= v_{\rm th} \sigma \num{\alpha \mathchar`- \ce{H2}}n_{\rm gr}, \\
%R_{\rm thdes}(\alpha \mathchar`- \ce{H2}) &= ,
%\end{align}
%where $v_{\rm th}$ is the thermal velocity, $\sigma$ is the cross-section of a dust grain, $n_{\rm gr}$ is the number density of dust grains per unit gas volume,
%and $k_{\rm thdes}$ is the thermal desorption rate (s$^{-1}$).

The first terms in Eqs. \ref{eq:gas_h2} and \ref{eq:cov_h2} represent adsorption to the surface, where $\sigma_{\rm gr}$ is the cross section of a dust grain.
The factor $1-\Theta$ or $1-\theta$ is included, assuming that only one molecule adsorbs per binding site.
%Then $S(1-\Theta)$ represents the sticking probability of \ce{H2} to the grain surface which is covered by water and \ce{H2} in our models.
Then the maximum value of $\Theta$ is unity, and the formation of \ce{H2} multilayers does not occur in our models.
The second term in Eqs. \ref{eq:gas_h2} and \ref{eq:cov_h2} represent thermal desorption.
The third term in Eq. \ref{eq:cov_h2} represents thermal hopping from site $i$ to site $j$, while the fourth term represents the reverse process, where 
$k_{\rm hop}$ is the rate constant of the thermal hopping.
%The hopping activation energy in our models is discussed later.
The fifth and sixth terms are for ortho-para conversion on surfaces.
Initially, all \ce{H2} are assumed to be present in the gas phase with the $\op{H2}$ of three (i.e., the statistical value).

The sticking probability of \ce{H2} is taken from \citet{he16}, which was experimentally measured for a non-porous amorphous solid water (ASW) surface.
The experimentally measured sticking probability would be considered as the surface averaged value, while the sticking probability for each site could depend on the energy depth of each site.
However, according to the molecular dynamics simulations of the \ce{H2} sticking on an ASW surface, \ce{H2} molecules visit several binding sites before remaining bound in one \citep{molpeceres20}.
Then, the use of the surface averaged sticking probability for Eqs. \ref{eq:gas_h2} and \ref{eq:cov_h2} would be acceptable in this work.
As the binding energy distribution of p-\ce{H2}, we adopt the experimentally determined binding energy distribution of \ce{D2} on non-porous ASW, considering the zero-point energy difference between \ce{D2} and \ce{H2} of $\sim$37 K \citep{he14,amiaud15}.
The binding energy of o-\ce{H2} at each site is set to higher than that of p-\ce{H2} by $\deopg - \Delta E_{\rm op,\,s}$ (Section \ref{sec:energy_difference}). 
The hopping activation energy from site $i$ to site $j$ is given by \citep{cazaux17}
\begin{linenomath*}
\begin{align}
%\ehop (\edes \rightarrow \edess) = &f \times {\rm min}(\edes,\,\,\edess) \nonumber \\
%&+ {\rm max}(0,\,\,\edes - \edess),
\ehop (i \rightarrow j) = &f \times {\rm min}(\edes(i),\,\,\edes(j)) \nonumber \\
&+ {\rm max}(0,\,\,\edes(i) - \edes(j)),
\end{align}
\end{linenomath*}
The hopping-to-binding energy ratio ($f$) for \ce{H2} is poorly constrained.
Laboratory studies found that the value of $f$ ranges from 0.2 to 0.7, depending on the species on ASW \citep{kouchi21,furuya22}.
The NSC on grains affects the OPR of overall (gas + ice) \ce{H2} more efficiently with lowering the hopping activation energy \citep{furuya19}.
Here we assume a relatively conservative value, $f = 0.65$.
The NSC timescale of \ce{H2} ($\tau^{\rm surf}_{\rm conv}$) is taken from \citet{ueta16}, who measured the timescale for ASW by laboratory experiments.
From $\tau^{\rm surf}_{\rm conv}$, the rate constant of the conversion from adsorbed o-\ce{H2} to adsorbed p-\ce{H2} ($k^{\rm surf}_{\rm op}$) and that of the reverse process ($k^{\rm surf}_{\rm po}$) can be deduced to be \citep{bron16}
\begin{linenomath*}
\begin{align}
k^{\rm surf}_{\rm op} &= 1/(\tau^{\rm surf}_{\rm conv}(1+\gamma_s)), \\
k^{\rm surf}_{\rm po} &= \gamma_s/(\tau^{\rm surf}_{\rm conv}(1+\gamma_s)).
\end{align}
\end{linenomath*}

Initially, all \ce{H2} is assumed to be in the gas phase with
an OPR of three.
We confirmed that the initial value of the \ce{H2} OPR does not affect our conclusion in this section or the derived values of $\eta_{\rm op}$ and $\eta_{\rm po}$.
We run a small grid of pseudo-time-dependent models (i.e., the gas density and the temperature are fixed in each model), varying the \ce{H2} gas density from 10$^3$ to 10$^7$ cm$^{-3}$ and the temperature from 10 to 20 K.
Gas and surface temperatures are assumed to be the same.
For each pseudo-time-dependent model, we consider the three case where $\Delta E_{\rm op,\,s}$ = $\Delta E_{\rm op,\,2Drot}$ (85 K), $\deops$ = 120 K, and $\Delta E_{\rm op,\,s}$ = $\deopg$ (170.5 K).
%In \citet{furuya19}, $\Delta E_{\rm op,\,s}$ was set to be the same as $E_{\rm op,\,3Drot}$, assuming that binding energy distributions for o-\ce{H2} and p-\ce{H2} were the same.

\subsubsection{Results}
\begin{figure*}[ht!]
\plotone{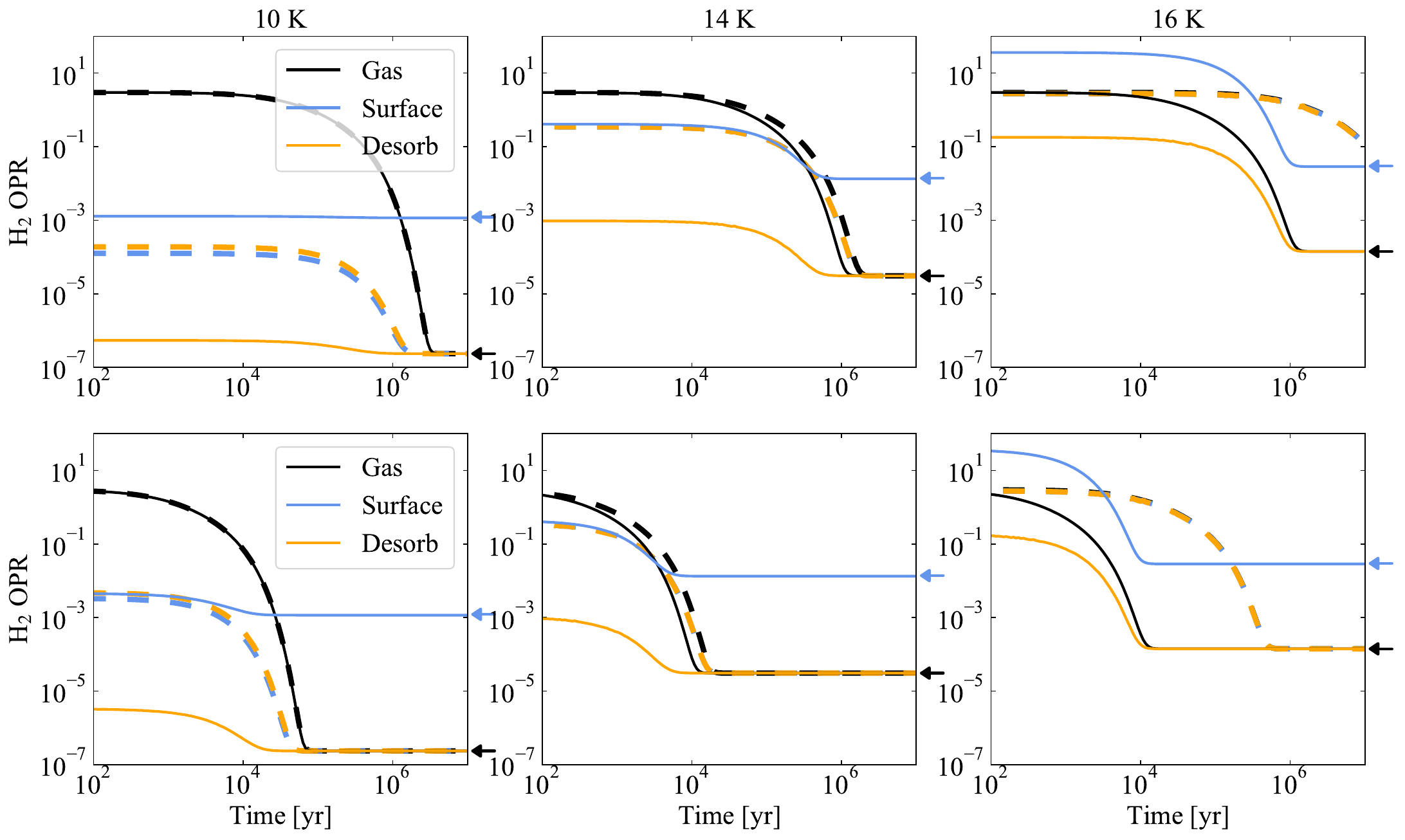}
\caption{Temporal evolution of the \ce{H2} OPR in the gas phase, on the surface, and in desorbing gas from the surface. Dashed lines show the model with $\deops = 170.5$ K, while solid lines show the model with $\deops = 85$ K.
The \ce{H2} gas density is 10$^4$ cm$^{-3}$ in top panels, while it is 10$^6$ cm$^{-3}$ in lower panels.
The temperature is from 10 K, 14 K, and 16 K in the left, middle, and right panels, respectively. 
Note that in the model with  $\deops = 170.5$ K, the OPRs on the surface and in the desorbing gas are similar, 
and the lines for them are partially overlapping in the figure. 
Arrows on the right margin indicate $6\exp(-170.5/T)$ (black) and $\gamma_s$ (blue).
}
\label{fig:delta_eop}
\end{figure*}

Figure \ref{fig:delta_eop} shows the temporal evolution of the \ce{H2} OPR in the gas phase, on the surface, and in desorbing gas from the surface (i.e., the ratio of the thermal desorption rate of o-\ce{H2} to that of p-\ce{H2}). The dashed lines represent the model with $\deops = 170.5$ K, while the solid lines represent the model with $\deops = 85$ K.
At 10 K, the evolution of the \ce{H2} OPR in the gas phase is almost identical regardless of the value of $\deops$.
The model with $\deops = 85$ K shows higher \ce{H2} OPR on the surface compared to the model with $\deops = 170.5$ K.
In contrast, the \ce{H2} OPR in the desorbing gas is lower in the model with $\deops = 85$ K than in the model with $\deops = 170.5$ K, because the binding energy for $\ohh$ at a given site is 85 K higher than that for $\phh$, leading to the lower desorption rate of $\ohh$ in the former model.
At $\gtrsim$10$^6$ yr, the \ce{H2} OPR of desorbing gas and in the gas phase become $\sim$6$\exp(-170.5/T)$ in both two models rather than $\gamma_s = 6\exp(\deops/T)$ or $\gamma_g = 9\exp(\deopg/T)$.

%On the other hand, the \ce{H2} OPR on the surface depends on $\deops$;  when $\deops$ = 170.5 K (85 K), the \ce{H2} OPR on the surface is  ($9\exp(-85/10)\sim 1.8\times10^{-3}$) at $\gtrsim$10$^6$ yr.
%How the different \ce{H2} OPR on the surface reflects that in the desorbing gas depends on time.
%At $\lesssim4\times10^5$ yr, the \ce{H2} OPR in the desorbing gas is different depending on $\deops$, reflecting the different \ce{H2} OPR on the surface.
%On the other hand, at the later times, the different \ce{H2} OPR on the surface is not reflected in the desorbing gas, the \ce{H2} OPR in the desorbing gas does not reflect 
%This difference comes from the fact that \ce{H2} coverage reaches almost the steady state, but the coverage of $\ohh$ evolves slowly because the deepest potential sites are occupied by $\phh$.
%Indeed, when we use the lower value for $f$, which increases the hopping rate, the \ce{H2} OPR in the desorbing gas in the two models becomes the same after XX yr.

At $\ge$14 K, unlike at 10 K, the temporal evolution of the \ce{H2} OPR in the gas phase depends on the value of $\deops$;
the timescale of the decrease of the \ce{H2} OPR is shorter in the model with $\deops = 85$ K than in the model with $\deops=170.5$ K.
This is becuase the higher binding energy of $\ohh$ in the model with $\deops = 85$ K results in longer residence time on the surface, which in turn leads to more efficient NSC than in the model with $\deops = 170.5$ K.
On the other hand, the steady-state value of the \ce{H2} OPR in the gas phase does not depend on $\deops$ and is given by 6$\exp(-170.5/T)$ as in the case of the temperature of 10 K.
We confirmed that this conclusion holds even when $\deops$ depends on sites by running additional models in which $\deops$ for a site with the binding energy of $\edes$ is given by $\Delta E_{\rm op,\,2Drot}\times(\edes/\edes_{,\,\rm max})$, where $\edes_{,\,\rm max}$ is the maximum binding energy.
These results indicate that when $\deops \neq \deopg$, instead of Eq. \ref{eq:op_surf3}, the slightly modified one should be used:
\begin{linenomath*}
\begin{align}
\frac{\eta_{\rm op}}{\eta_{\rm po}} = 6\exp\left(-\frac{\deopg}{T_{\rm s}}\right). \label{eq:op_surf4}
\end{align}
\end{linenomath*}

%The temperature above which the gas-phase \ce{H2} OPR evolution depends on $\deops$ can be evaluated as follows.
%By considering the balance between the spin conversion rate of o-\ce{H2} to p-\ce{H2} ($k^{surf}_{op}$) and the thermal desorption rate from the deepest site for o-\ce{H2} in the model with $\deops = 170.5$ K ($\nu \exp(-550/T_s)$), we obtain the critical temperature of $\sim$16 K.
%Then, at $\gtrsim$16 K, the desorption timescale of o-\ce{H2} is too short for the NSC on the surface unless $\deops < E_{\rm op,\,3Drot}$.

\begin{figure}[ht!]
\epsscale{0.6}
\plotone{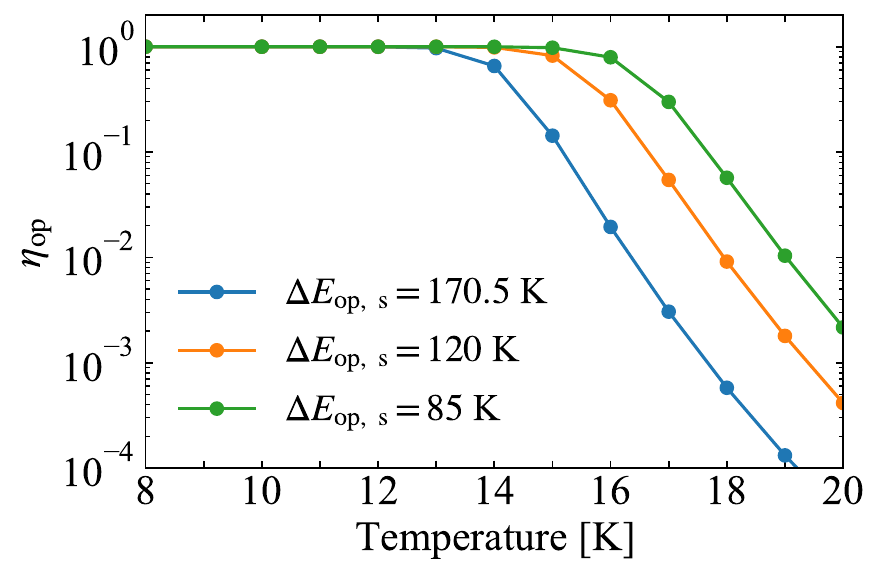}
\caption{The parameter $\eta_{\rm op}$ as functions of the temperature in the case when $\deops = 170.5$ K (blue), and $\deops = 120$ K (orange), and $\deops = 120$ K (green).
As the density dependence of $\eta_{\rm op}$ is very weak, the values only at the \ce{H2} density of 10$^4$ cm$^{-3}$ are plotted here.
}
\label{fig:eta}
\end{figure}

Under the adsorption-desorption equilibrium which is established within $\sim$1 yr ($10^4$ cm$^{-3}$/$n$(\ce{H2})), the parameter $\eta_{\rm op}$ is derived from the desorption rates of o-\ce{H2} and p-\ce{H2} in the numerical simulations \citep[see Sec. 3.4 of][]{tsuge21a}.
The resulting $\eta_{\rm op}$ at the density of 10$^4$ cm$^{-3}$ is shown in Figure \ref{fig:eta}.
Note that although $\deops$ does not explicitly included in Eqs. \ref{eq:op_surf1} and \ref{eq:op_surf2},
$\eta_{\rm op}$ (and $\eta_{\rm po}$) do depend on the value of $\deops$.

\epsscale{1.0}

Finally, we confirmed that numerically integrating the following equations for the temporal evolution of o-\ce{H2} and p-\ce{H2}, using the derived $\eta_{\rm op}$ and $\eta_{\rm po}$ (see Eq. \ref{eq:op_surf4}), successfully reproduces the temporal evolution of the \ce{H2} OPR in the gas phase obtained by solving Eqs. \ref{eq:gas_h2} and \ref{eq:cov_h2}:
\begin{linenomath*}
\begin{align}
\frac{dn({{\rm o} \mathchar`- \ce{H2}})}{dt} &= -R_{\rm op, \,s} + R_{\rm po, \,s}, \\
\frac{dn({{\rm p} \mathchar`- \ce{H2}})}{dt} &= -R_{\rm po, \,s} + R_{\rm op, \,s}.
\end{align}
\end{linenomath*}
Then, we conclude that in general cases, once $\eta_{\rm op}$ is given, the rates for the temporal evolution of the \ce{H2} OPR through NSC on surfaces are given by Eqs. \ref{eq:op_surf1}, \ref{eq:op_surf2} and \ref{eq:op_surf4}, which can be included in any astrochemical models based on the rate-equation approach.

%When $\deops \lesssim E_{\rm op,\,3Drot}$, on the other hand, the desorption timescale of o-\ce{H2} in the deepest sites can be long enough for the NSC on the surface. 
%make a plot for multiple coverage_dist.d files.

%We conclude that the actual value of $\deops$ does not matter for the steady state OPR of the bulk \ce{H2}, while the timescale for gas-phase \ce{H2} OPR evolution via NSC on surfaces depends on $\deops$.
%In other words, Eq. \ref{eq:op_surf3} holds regardless of the value of $\deops$, while $\eta_{op}$ and $\eta_{po}$ depend on $\deops$.

\section{Timescale of the \ce{H2} OPR evolution and its steady state value in the ISM} \label{sec:steady_state}
Here we discuss the timescale of the \ce{H2} OPR evolution and its steady state value ($\opst$) in the ISM, which are determined by the competition among the \ce{H2} formation, the NSC via the gas-phase reactions, and the NSC on grain surfaces.
Assuming all atomic H adsorbed onto dust grains are consumed by the \ce{H2} formation, the \ce{H2} formation timescale is given by
\begin{linenomath*}
\begin{align}
%R_{\rm \ce{H2}} = 0.5S\sigma_{\rm gr} v_{\rm th}n_{\rm gr}n(\ce{H}),
\tau_{\rm \ce{H2}} = n({\rm H_2})/(0.5S_{\rm H_2}R_{\rm col}(\ce{H})),
\end{align}
\end{linenomath*}
where $R_{\rm col}(\ce{H})$ is the collisional rate (cm$^{-3}$ s$^{-1}$) of atomic H to dust grains.
%where $S$ is the sticking probability, $\sigma_{\rm gr}$ is the geometrical cross-section of a dust grain, $v_{\rm th}$ is the thermal velocity of atomic H, and 
%$n_{\rm gr}$ is the number density of dust grain in a unit gas volume.
%$n(X)$ indicates the number density of species $X$ throughout this paper.
The NSC timescale via the gas-phase proton exchange reactions is given by
\begin{linenomath*}
\begin{align}
%R_{\rm op,\,g} &= k_{\rm op,\,g}n(\ce{H3+})n(\ohh), \,\,\,\, \tau_{\rm op,\,g} = n(\ohh)/R_{\rm op,\,g}, \\
%R_{\rm po,\,g} &= k_{\rm op,\,g}\gamma_{\rm g}n(\ce{H3+})n(\phh), \,\,\,\, \tau_{\rm op,\,g} = n(\phh)/R_{\rm po,\,g},
\tau_{\rm op,\,g} &= 1/(k_{\rm op,\,g}n(\ce{H3+})),  \\
\tau_{\rm po,\,g} &= 1/(k_{\rm op,\,g}\gamma_{\rm g}n(\ce{H3+})),
\end{align}
\end{linenomath*}
where $k_{\rm op,\,g}$ is the rate constant of the proton exchange reactions in the gas-phase \citep{honvault11,honvault12,hilyblant18}.
Here we consider the proton exchange reactions with \ce{H3+} only because the exchange reactions with \ce{H3+} are more important than those with \ce{H+} in the dense ISM \citep{furuya15}.
We estimate the abundance of \ce{H3+} using the analytical method introduced by \citet{aikawa15} (see their Sec. 4), but neglecting the presence of \ce{N2H+} and recombination of ions with charged grains.
With this method, we can self-consistently estimate the abundance of \ce{H3+}, \ce{HCO+}, and electrons for given parameters, including the cosmic-ray ionization rate of \ce{H2} ($\zeta$), the CO abundance, temperature, and the gas density.
%Assuming \ce{H3+} is the dominant positive charge carrier, the \ce{H3+} abundance is estimated by $\sqrt{\xi/(k_{\rm e}n_{\rm gas})}$, where $\xi$ is the cosmic-ray ionization rate of \ce{H2}, and $k_{\rm e}$ is the rate coefficient for dissociative recombination of \ce{H3+}.
Finally, the timescale for the NSC on surfaces is given by
\begin{linenomath*}
\begin{align}
\tau_{\rm op,\,s} &= n(\ohh)/(\eta_{\rm op}S_{\ce{H2}}(1-\Theta)R_{\rm col}(\ohh)),  \\
\tau_{\rm po,\,s} &= n(\phh)/(\eta_{\rm po}S_{\ce{H2}}(1-\Theta)R_{\rm col}(\phh)).
\end{align}
\end{linenomath*}
%($\tau_{\rm \ce{H2}} = n({\rm H_2})/R_{\rm H_2}$)

\begin{figure*}[ht!]
\plotone{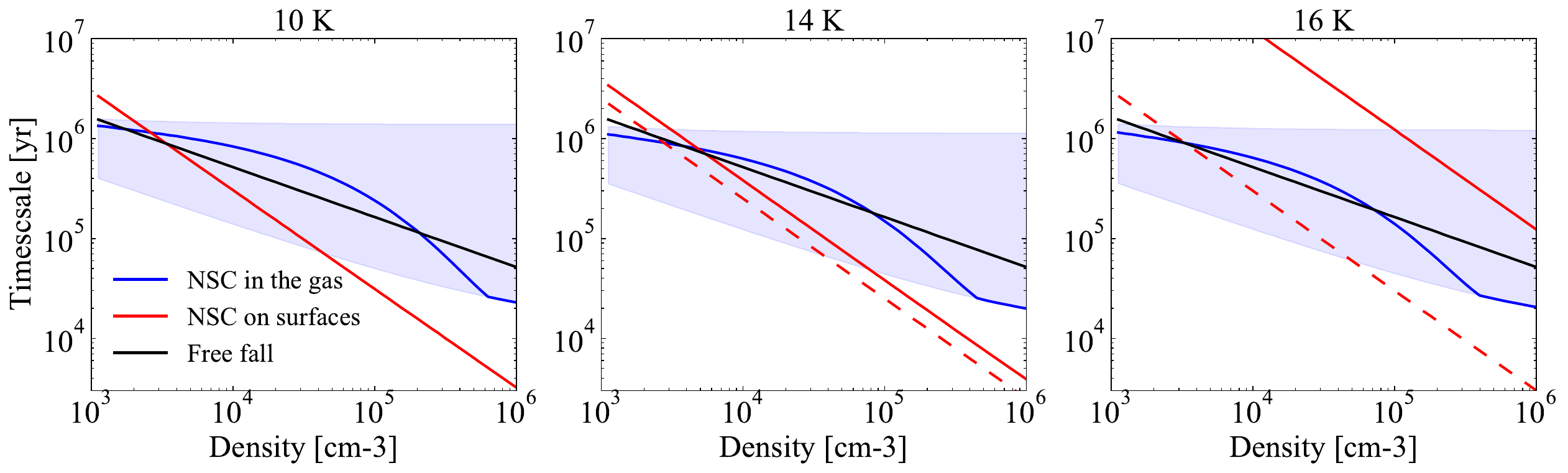}
\caption{Conversion timescale of o-\ce{H2} to p-\ce{H2} via the NSC on grain surfaces when $\deops=170.5$ K (red solid lines) and when $\deops=85$ K (red dashed lines) as functions of the gas density.
Black solid line indicates the free-fall timescale ($\tau_{\rm ff}$). 
The blue area indicates the conversion timescale of o-\ce{H2} to p-\ce{H2} via the gas-phase proton exchange reaction with the upper and lower ends corresponding to the CO abundance of 10$^{-4}$ and 10$^{-6}$, respectively. The blue solid line indicates the case when the CO abundance is given by 10$^{-4}\exp(-\tau_{\rm ff}/\tau_{\rm freeze})$, where $\tau_{\rm freeze}$ is the freeze-out timescale of CO onto dust grains, with the minimum CO abundance of 10$^{-6}$.
Gas and dust temperatures are the same and assumed to be 10 K, 14 K, and 16 K in the left, middle, and right panels, respectively.}
\label{fig:h2opr_timescale}
\end{figure*}

Figure \ref{fig:h2opr_timescale} shows $\tau_{\rm op,\,g}$, $\tau_{\rm op,\,s}$, and the free-fall timescale ($\tau_{\rm ff}$) the last of which is a measure of the dynamical timescale of star formation, representing the shortest possible timescale for gravitational collapse.
In the figure, it is assumed that the cosmic-ray ionization rate of \ce{H2} is 10$^{-17}$ s$^{-1}$, the dust-to-gas mass ratio is 0.01, and the dust grain radius is 0.1 $\mu$m.
It is worth noting the different density dependence of the timescales.
Since $R_{\rm col}(\ce{H}) \propto n_{\rm gr}n(\ce{H})$, and since $n({\ce{H}})$ is $\sim$1($\zeta/10^{-17}$ s$^{-1}$) cm$^{-3}$, regardless of the gas density, in regions where the UV radiation is well-shielded and hydrogen is mostly locked up in \ce{H2} \citep[e.g.,][]{goldsmith05}, $\tau_{\rm H_2}$ does not depend on the gas density (denoted as $n_{\rm gas}$), and is $\sim$3$\times 10^9$ yr, which is much longer than the other timescales (thus not shown in Fig. \ref{fig:h2opr_timescale}). 
$\tau_{\rm op,\,s}$ is proportional to $\sim n_{\rm gas}^{-1}$, as the density dependencies of $\eta$ and $\Theta$ are relatively weak \citep{furuya19}. 
$\tau_{\rm op,\,g}$ is proportional to $\sim n_{\rm gas}^{-0.5}$ when \ce{H3+} is the dominant positive charge carrier (i.e., when the CO abundance is low; see the lower end of the blue area in Fig. \ref{fig:h2opr_timescale}), because the number density of major ions (i.e., \ce{H3+}) are proportional to $\sim n_{\rm gas}^{0.5}$ \citep[e.g.,][]{tielens05}.
When the CO abundance is high (the upper end of the blue area), \ce{HCO+} rather than \ce{H3+} is the dominant charge carrier and the number density of \ce{H3+} does not depend on $n_{\rm gas}$ \citep[see Eq. 23 of ][]{aikawa15}, and thus $\tau_{\rm op,\,g}$ neither.
$\tau_{\rm ff}$ is scaled with $n_{\rm gas}^{-0.5}$ \citep{spitzer78}.
Taken together, the NSC on surfaces has the strongest density dependence, and in regions with the density of $\gtrsim$10$^4$ cm$^{-3}$, the NSC on surfaces plays a major role in the evolution of the \ce{H2} OPR.
The latter argument is consistent with the hydrodynamical simulations of collapsing filaments coupled with chemistry by \citet{lupi21}, which showed the NSC in the gas-phase is more important than that on grains in the filaments where the gas density is mostly $\lesssim$10$^4$ cm$^{-3}$.

The \ce{H2} OPR would reach the steady state at a density higher than 10$^5$ cm$^{-3}$, as the timescale of NSC is much shorter than the free-fall timescale.
If the dynamical evolution proceeds more slowly than the free-fall timescale due to support from thermal pressure, magnetic fields, or other effects, this threshold density (10$^5$ cm$^{-3}$) becomes lower.
While $\tau_{\rm op,\,g}$ and $\tau_{\rm ff}$ do not strongly depend on the temperature, $\tau_{\rm op,\,s}$ becomes larger with increasing the temperature.
When $\deops = 170.5$ K, the NSC on grains is negligible compared to the NSC in the gas phase at $\gtrsim$16 K.
On the other hand, when $\deops = 85$ K, even at 16 K, the NSC on grains is more efficient than that in the gas phase.
At 20 K, the NSC on grains becomes negligible compared to the NSC in the gas phase even when $\deops = 85$ K.
Then, the actual value of $\deops$ becomes critical at slightly warmer temperatures than the typical temperature of molecular clouds ($\sim$10 K).

Note that in this section, we fixed the cosmic ionization rate of \ce{H2} to be $\zeta = 10^{-17}$ s$^{-1}$. As the abundances of ion species increases with increasing $\zeta$, the timescale for the NSC in the gas phase becomes shorter with increasing $\zeta$. For the galactic center environments, where $\zeta$ is $10^{-14}$ s$^{-1}$ or even higher \citep{LePetit16}, the threshold density at which NSC on grains is more efficient than that in the gas phase would increase to above $\sim$10$^5$-10$^6$ cm$^{-3}$. In addition, the timescale of the NSC in the gas phase is sufficiently short compared to the free-fall timescale. Then, in that environment, the role of NSC on grains would be minor.

\begin{figure*}[ht!]
\plotone{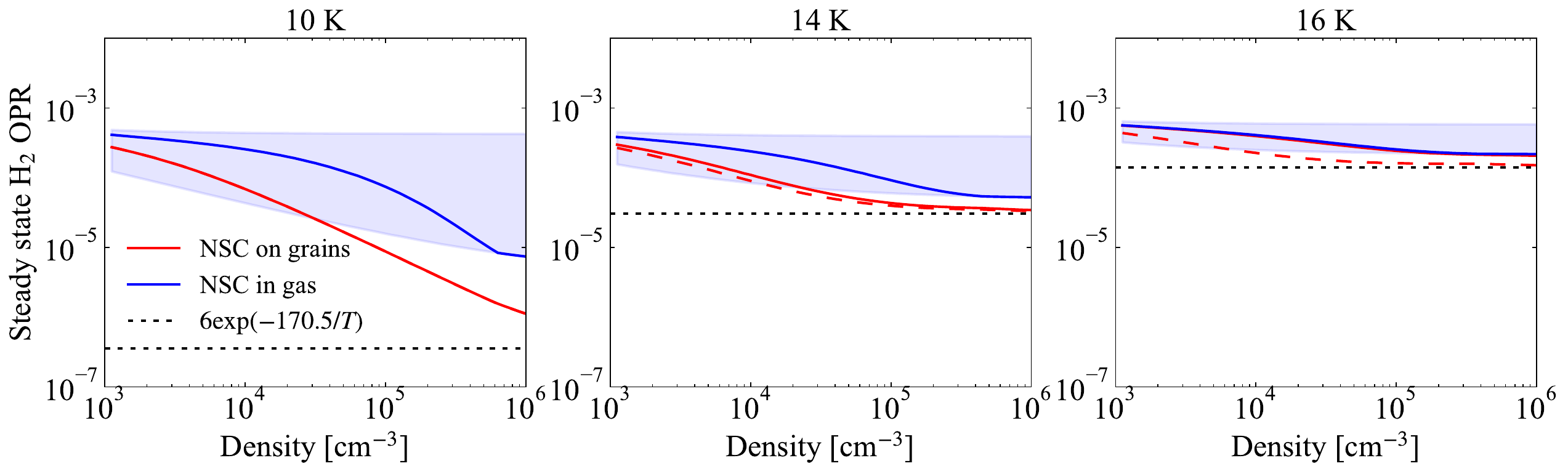}
\caption{A steady state value of \ce{H2} OPR as a function of gas density. Gas and dust temperatures are the same and assumed to be 10 K, 14 K, and 16 K in the left, middle, and right panels, respectively. Red solid lines indicate the case when both the NSC on grains and that in the gas phase are included.
The blue area indicates the case when only the NSC in the gas-phase is considered, with the upper and lower ends corresponding to the CO abundance of 10$^{-4}$ and 10$^{-6}$, respectively. The blue solid line indicates the case when the CO abundance is given by 10$^{-4}\exp(-\tau_{\rm ff}/\tau_{\rm freeze})$, where $\tau_{\rm freeze}$ is the freeze-out timescale of CO onto dust grains, with the minimum CO abundance of 10$^{-6}$.
Black dotted lines correspond to $6\exp(-170.5/T)$.}
\label{fig:h2opr_st}
\end{figure*}

In the case where only the \ce{H2} formation and the proton exchange reactions in the gas phase are considered, the analytical expression of $\opst$ was derived by \citet{furuya15} (see their Appendix A).
It is straightforward to include the contribution of the NSC on solid surfaces, and one can obtain
\begin{linenomath*}
\begin{align}
\opst &= \frac{\tau_{\rm op}/\tau_{\rm po} + b_0\tau_{\rm op}/\tau_{\rm \ce{H2}}}{1+(1-b_0)\tau_{\rm op}/\tau_{\rm \ce{H2}}}, \label{eq:opst}
\end{align}
\end{linenomath*}
where $b_0$ is the branching ratio to form o-\ce{H2} upon the \ce{H2} formation \citep[0.75;][]{watanabe10}.
The conversion timescale of o-\ce{H2} to p-\ce{H2} ($\tau_{\rm op}$) is defined as
\begin{linenomath*}
\begin{align}
\frac{1}{\tau_{\rm op}} = \frac{1}{\tau_{\rm op,\,g}} + \frac{1}{\tau_{\rm op,\,s}}. 
\end{align}
\end{linenomath*}
%where $\tau^g_{\rm op}$ and $\tau^s_{\rm op}$ are the conversion timescale of o-\ce{H2} to p-\ce{H2} via the gas-phase proton exchange reactions and that via the NSC on solid surfaces, respectively.
The NSC timescale of p-\ce{H2} to o-\ce{H2} ($\tau_{\rm po}$) is defined in a similar way.
%When the NSC on dust grains is neglected (i.e., $1/\tau^s_{\rm op} = 0$), Eq. \ref{eq:opst} is reduced to the steady-state \ce{H2} OPR derived by \citet{furuya15}.
%\tau_{\rm \ce{H2}}$ is the formation timescale of \ce{H2} on grain surfaces, that is defined as $n({\rm H_2})/R_{\rm H_2}$.
%The NSC timescale in the gas phase and on grain surfaces are defined as $\tau^g_{\rm op} = n(\ohh)/R_{\rm op}^{\rm gas}$ and $\tau^s_{\rm op} = n(\ohh)/R_{\rm op}^{\rm surf}$, respectively.
Although $\tau_{\rm H_2}$ is much larger than the NSC timescales, the former is still relevant to the steady state value of \ce{H2} OPR.
This is because the statistical value of three is much larger than $\gamma_g$ at low temperatures ($\sim$10 K).
Figure \ref{fig:h2opr_st} shows the steady state value of \ce{H2} OPR predicted by Eq. \ref{eq:opst} varying gas density and temperature.
The steady-state \ce{H2} OPR becomes lower than that without the NSC on grain surfaces, in particular at the temperature of 10 K and the density of $>$10$^4$ cm$^{-3}$.
The difference due to the NSC on grains would be minor for deuterium fractionation chemistry, because the steady state \ce{H2} OPR is lower than 10$^{-3}$ even in the case without the NSC on grains (see Fig. \ref{fig:analytical}).
Thus, the primary role of NSC on grains is to shorten the \ce{H2} OPR timescale at the density of $>$10$^{4}$ cm$^{-3}$ and the temperature of $\lesssim$14-16 K (depending on the value of $\deops$), rather than to lower the steady-state value of the \ce{H2} OPR.

%With higher density, the impact of the NSC becomes more significant.

\section{Modeling of deuterium fractionation chemistry with the \ce{H2} OPR in prestellar cores} \label{sec:model}

%\subsection{Physical model}
%We consider two different types of physical models.
%In one model, we adopt a pseudo-time-dependent model, where physical conditions are constant with time.
%The density is set to either 10$^4$ cm$^{-3}$ or 10$^7$ cm$^{-3}$. 
%Gas and dust temperatures are set to be 10 K, while the visual extinction is set to be 10 mag.

\subsection{Astrochemical model description}
Here we discuss the impact of the NSC on grain surfaces on the deuterium fractionation in a collapsing prestellar core.
To this end, we use a gas-ice astrochemical code, Rokko, presented in \citet{furuya15}, which considers mono-, doubly, and triply deuterated species and nuclear spin states of \ce{H2}, \ce{H3+}, and their isotopologues.
We adopt the three-phase model, which includes chemical species in the gas phase, on the surface of the ice, and in the bulk ice mantle, assuming that the bulk mantle has uniform chemical composition \citep{hasegawa93}.
We consider gas-phase reactions, interactions between gas and (icy) grain surfaces, and surface reactions as chemical processes.
As non-thermal desorption mechanisms, which are more important than thermal desorption at low temperatures ($\sim$10 K), photodesorption, chemical desorption, and whole grain heating by cosmic rays are taken into account.
See \citet{furuya15} and references therein for more details.
%The desorption rates by the whole grain heating by cosmic-rays are calculated according to \citet{sipila21}.

%Our chemical network is originally based on that in \citet{garrod06} and was extended to include singly, doubly, and triply deuterated species and the nuclear spin states of \ce{H2}, \ce{H3+}, and their isotopologues \citep{aikawa12,coutens14}.

There are some updates from \citet{furuya15}.
We use the species-to-species rate coefficients for the \ce{H2}-\ce{H3+} reaction systems from \citet{hilyblant18}, assuming local thermodynamic equilibrium for the rotationally excited states within each nuclear spin state \citep[see also][]{sipila17}.
In \citet{furuya15}, the rate coefficients for the \ce{H2}–\ce{H3+} reaction systems were adopted from \citet{hugo09}, assuming that the species in each nuclear spin state are in their rotational ground states.
The rate coefficients for the isotope exchange reactions of D atoms with \ce{H3+} or \ce{H2D+} are set to be 10$^{-12}$ s$^{-1}$, which corresponds to the upper limit value obtained by experiments and quantum chemical calculations \citep{hillenbrand19}.
In \citet{furuya15}, their rate coefficients were set to be three orders of magnitude higher \citep[10$^{-9}$ cm$^{-3}$;][]{millar89} than the values used in this work.
%As discussed in \citep{sipila17},  
As in \citet{furuya15}, the rate coefficient for the proton exchange reaction of \ce{H2} with \ce{H+} is taken from \citet{honvault11,honvault12}, and the OPR upon the \ce{H2} formation is set to the nuclear-spin statistical value of 3 \citep{watanabe10,tsuge19}.
The NSC of \ce{H2} on surfaces is newly considered with Eqs. \ref{eq:op_surf1}, \ref{eq:op_surf2}, and \ref{eq:op_surf4}.
The parameter $\eta$ is taken from the simulation results presented in Section \ref{sec:detail_model}.
We adopt a quasi-steady state treatment of the surface coverage of \ce{H2}, assuming the adsorption-desorption equilibrium \citep{furuya24}.
We also adopt the same treatment for HD and \ce{D2} on grain surfaces as \ce{H2}.

The elemental abundances are taken from \citet{aikawa99}.
The D/H elemental abundance ratio is set to be $1.5\times10^{-5}$
\citep{linsky03}.
Initially, hydrogen and deuterium are present as \ce{H2} and HD, respectively.
The other elements are initially either neutral atoms or atomic ions, depending on their ionization energy.
Astrochemical modeling suggests that the \ce{H2} OPR is much lower than unity in molecular clouds and prestellar cores \citep{furuya15,lupi21}.
There are some indirect estimates of the \ce{H2} OPR in prestellar cores from the observations of deuterated molecules, suggesting that the \ce{H2} OPR is $\sim$10$^{-2}$ for Barnard 68 \citep{maret07} and $\sim$10$^{-2}$ or even higher for L183 \citep{pagani09,pagani13}.
Here the initial \ce{H2} OPR is treated as a free parameter, and three different values ($10^{-3}$, $10^{-2}$, and $0.1$) are considered.

We simulate the molecular evolution, assuming the physical conditions evolve with time on a timescale of free-fall collapse, mimicking a collapsing prestellar core \citep{brown89}.
The initial and final gas densities are set to $2\times10^3$ cm$^{-3}$ and 10$^7$ cm$^{-3}$, respectively.
%We assume that the core is well-shielded from the interstellar UV radiation.
The visual extinction ($A_V$) is assumed to increase following $n_{\rm gas}^{2/3}$ \citep{garrod11} with the initial $A_V$ of 2 mag.
The dust temperature is calculated using Equation (8) in \citet{hocuk17} with the floor value of 10 K.
The gas temperature is assumed to be the same as the dust temperature.
Note that, as the maximum temperature in the physical model is $\sim$13 K, the efficiency of the NSC on grains does not depend on $\deops$ (see Figs. \ref{fig:h2opr_timescale} and \ref{fig:h2opr_st}) in this simulation setup.
Cosmic-ray ionization rate of \ce{H2} is set to be $1.3\times10^{-17}$ s$^{-1}$.

%In the kinetic Monte-Carlo simulations, it is often assumed that \ce{H2} is allowed to hop or adsorb on another \ce{H2} without the barrier for the encounter of two \ce{H2} molecules \citep{cuppen07}.
%According to our quantum chemistry calculations presented in Sect. \ref{sec:quantum}, the assumption is not justified.

%\section{Results} \label{sec:result}

%\subsection{Pseudo-time-dependent model}
%\begin{figure*}[ht!]
%\plotone{DH_H3p.eps}
%\caption{Left) Temporal evolution of the total abundance of \ce{H3+} isotopologues (top panels), the abundance ratios of \ce{H3+} isotopologues (middle panels), and the \ce{H2} OPR (bottom panels) in pseudo-time dependent models with the gas density of 10$^4$ cm$^{-3}$ (left panels) and 10$^7$ cm$^{-3}$ (right panels). Gas and dust temperatures are fixed to be 10 K. Solid, dashed, and dotted lines indicate the model with $\deops$ = 80 K, with $\deops$ = 170 K, and without spin conversion on the surface.}
%\label{fig:dh_pseudo}
%\end{figure*}
\subsection{Results}
\begin{figure*}[ht!]
\plotone{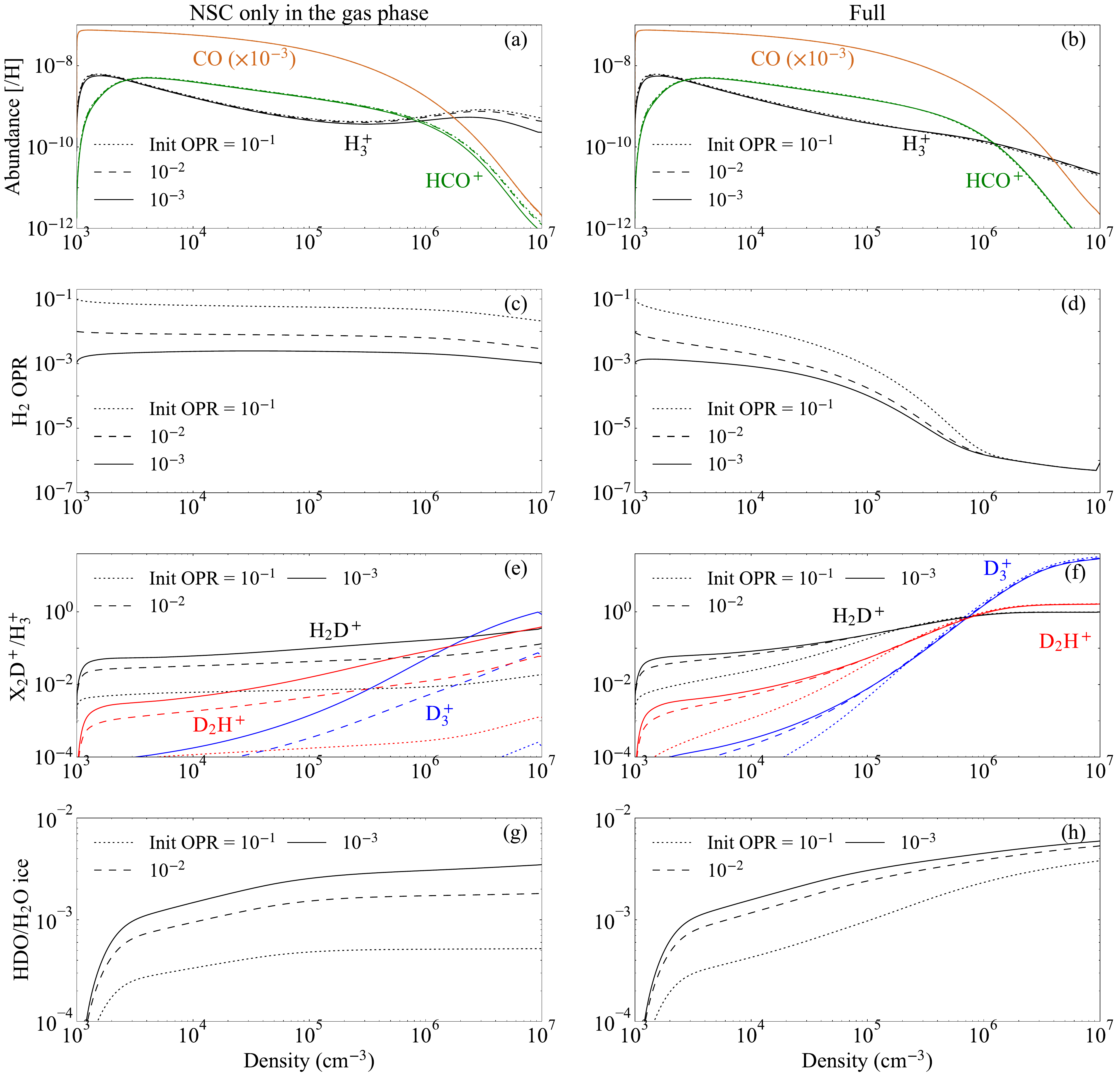}
\caption{
The evolution of the abundances and abundance ratios in the collapse model with (right panels) and without (left panels) the NSC on grains.
Solid, dashed, and dotted lines represent models with the initial \ce{H2} OPR of 10$^{-3}$. 10$^{-2}$, and 10$^{-1}$, respectively.
(a), (b) the abundances of selected species. (c), (d) the \ce{H2} OPR. (e), (f) the \ce{X2D+}/\ce{H3+} abundance ratio, where X represents H or D. (g), (h) the icy HDO/\ce{H2O} ratio.
In panels (a) and (b), the CO abundance is multiplied by 10$^{-3}$ to increase the visibility, the \ce{H3+} abundance is the sum of ortho and para states, and the models with the different initial \ce{H2} OPR are depicted with the same color, because it does not significantly affect the abundances of the selected species.
As the density increases with time on the free-fall timescale in the model, the horizontal axes can be read as time.}
\label{fig:collapse}
\end{figure*}

\begin{figure}[ht!]
\centering
\includegraphics[width=0.4\textwidth]{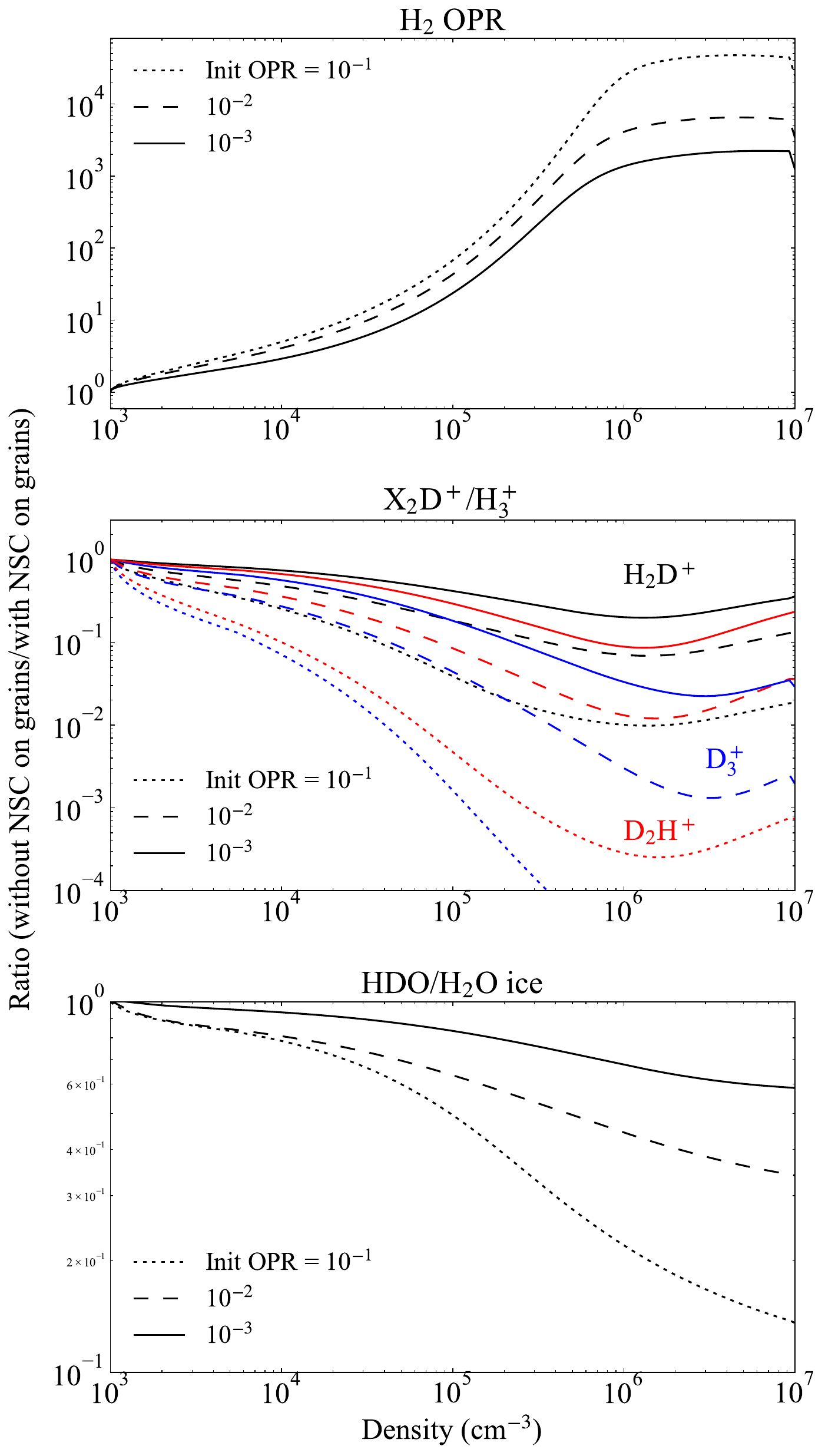}
\caption{
Ratio of the results obtained without and with NSC on grains in the collapse model: (top) \ce{H2} OPR, (middle) \ce{X2D+}/\ce{H3+}, where X is H or D, and (bottom) icy \ce{HDO}/\ce{H2O}.
Solid, dashed, and dotted lines represent models with the initial \ce{H2} OPR of 10$^{-3}$. 10$^{-2}$, and 10$^{-1}$, respectively.
}
\label{fig:collapse_ratio}
\end{figure}

%\begin{figure*}[ht!]
%\plotone{pseudo.pdf}
%\caption{Temporal evolution of the \ce{H2} OPR in the gas-phase and (top panels) the %\ce{H2D+} abundance (bottom panels) with respect to hydrogen nuclei in the pseudo-time dependent models. The gas density is 10$^4$ cm$^{-3}$. The gas and dust temperatures are 10 K (left panels), 14 K (middle panels), and 18 K (right panels). Solid lines and dashed lines show models with and without NSC on grains, respectively.}
%\label{fig:pseudo}
%\end{figure*}
%Firstly we consider pseudo-time-dependent models, where physical conditions are fixed with time.
%We consider three different temperatures (10 K, 14 K, and 18 K) and two different densities (10$^4$ cm$^{-3}$ and 10$^5$ cm$^{-3}$), varying in the treatment of the NSC on grains.
%The initial value for the \ce{H2} OPR is assumed to be three.
%Figure \ref{fig:pseudo} shows the temporal evolution of the \ce{H2} OPR in the gas phase and the \ce{H2D+} abundance with respect to hydrogen nuclei.
Figure \ref{fig:collapse} shows the abundances of selected species (panels a and b), the \ce{H2} OPR in the gas phase (panels c and d), and the \ce{X2D+}/\ce{H3+} ratio, where X is H or D (e and f), as functions of the gas density in the model with (left panels) and without NSC on grains (right panels).
Figure \ref{fig:collapse_ratio} shows ratios of the results obtained without and with NSC on grains for ease of comparison. 
In our models, the density increases with time on the free-fall timescale, so the horizontal axes can be read as time.
In the model without the NSC on grains, the \ce{H2} OPR and the \ce{X2D+}/\ce{H3+} ratios depend on the initial value of the \ce{H2} OPR even at the highest density (or the final simulation time),
because $\tau_{\rm op,\,g}$ is larger than $\tau_{\rm ff}$ until the CO freeze-out becomes significant (Fig. \ref{fig:h2opr_timescale}).
%The \ce{H2D+}/\ce{H3+} ratio at the highest density is $\sim$2$\times 10^{-2}$, $\sim$0.1, and $\sim$0.3 in the models with the initial \ce{H2} OPR of 10$^{-3}$, 10$^{-2}$, and 0.1, respectively. 
This trend is qualitatively consistent with the earlier modeling studies \citep[e.g.,][]{flower06}. 

In the case when the NSC on grains is included, the impact of the initial \ce{H2} OPR becomes smaller at the density higher than 10$^4$ cm$^{-3}$, because  $\tau_{\rm op,\,g}$ is shorter than $\tau_{\rm ff}$ (Fig. \ref{fig:h2opr_timescale}).
Regardless of the initial value, the \ce{H2} OPR becomes as low as $\sim$10$^{-4}$ and the \ce{H2D+}/\ce{H3+} ratio becomes $>$0.1 at the density of $\gtrsim$10$^5$ cm$^{-3}$.
At the highest density ($\gtrsim$10$^6$ cm$^{-3}$), the \ce{H2D+}/\ce{H3+} ratio is around unity and \ce{D3+} is the most abundant among the \ce{H3+} isotopologues \citep[e.g.,][]{wakelam04}.
The \ce{D2H+}/\ce{H3+} and \ce{D3+}/\ce{H3+} ratios at the highest density also do not depend on the initial \ce{H2} OPR.
The \ce{X2D+}/\ce{H3+} ratios in the model with NSC on grains are higher than those in the model without NSC on grains.
The impact is more significant for \ce{D2H+} and \ce{D3+} comapred to \ce{H2D+}, because the \ce{H2D+}/\ce{H3+} ratio saturates around unity due to the \ce{H2D+} destruction by HD (i.e., the \ce{D2H+} formation).
Therefore, including NSC on grains reduces the sensitivity of deuterium fractionation chemistry to the assumed initial \ce{H2} OPR, and enhances the deuterium fractionation at the high densities ($\gtrsim$10$^4$ cm$^{-3}$).
It should be noted, however, that the above statement is true for locally formed species, such as ion molecules, but not, e.g., water ice;
the icy \ce{HDO}/\ce{H2O} ratio (dashed lines in the bottom panels) still depends on the initial \ce{H2} OPR at the density higher than 10$^{5}$ cm$^{-3}$ even in the model with the NSC on grains, because water ice continuously forms and accumulates from the beginning to the end of the simulation.

\section{Discussion} \label{sec:discuss}
\subsection{The OPR of \ce{H2D+} in a protostellar envelope}
It has been proposed that the \ce{H2} OPR can serve as a chemical clock for star formation, since it is initially 3 upon the \ce{H2} formation, is expected to decrease monotonically over time in cold ($\sim$10 K) regions, and the timescale of NSC in the gas phase is long (Section \ref{sec:steady_state}).
However, as the NSC on grains can shorten the evolutionary timescale of the \ce{H2} OPR, it would be useful to reconsider its usefulness and limitations, particularly at the high-density regions ($\gtrsim$10$^4$ cm$^{-3}$).
Deuterium fractionation ratio, such as \ce{N2D+}/\ce{N2H+}, has often been used as an indirect tracer of the \ce{H2} OPR, because it is difficult to directly observe \ce{H2} in the cold gas ($\sim$10 K) \citep[e.g.,][]{pagani11}.
Here we focus on the \ce{H2D+} OPR, which has been proposed as the best tracer of the \ce{H2} OPR \citep{brunken14}.
Since the \ce{H2D+} OPR is set by the collision between \ce{H2D+} and \ce{H2}, the \ce{H2D+} OPR is given as a function of the \ce{H2} OPR and gas temperature \citep{gerlich02,brunken14}.
Then, if the gas temperature is given, the OPR of \ce{H2} and that of \ce{H2D+} have a one-to-one relation.
\citet{brunken14} detected o-\ce{H2D+} and p-\ce{H2D+} in the cold outer envelope of the Class 0 protostellar source IRAS 16293-2422.
They derived the \ce{H2D+} OPR of $ 0.065\pm0.019$ in the region with the gas temperature of 13-16 K, which corresponds to the \ce{H2} OPR of $\sim$2$\times10^{-4}$.
In order to interpret the low \ce{H2} OPR, they performed pseudo-time-dependent astrochemical simulations with the physical conditions appropriate for the cold outer envelope, assuming the initial \ce{H2} OPR of 10$^{-3}$.
They claimed that to reduce the \ce{H2D} OPR and \ce{H2} OPR to the level derived from the observations, it takes more than 1 Myr, which is longer than the typical lifetime of the Class 0 stage \citep{dunham14}.
Because their astrochemical model did not consider the NSC of \ce{H2} on grain surfaces, their estimated timescale should be regarded as the upper limit value.
It should be noted that the estimated timescale by \citet{brunken14} has been questioned;
\citet{harju17} re-estimated the timescale of 0.5 Myr by using a pseudo-time-dependent astrochemical model similar to that of \citet{brunken14}, but considered both the \ce{H2D+} and \ce{D2H+} data toward IRAS 16293-2422 in their analysis.
\citet{bovino21}, on the other hand, proposed even shorter timescale, 0.2 Myr, employing three-dimensional magneto-hydrodynamical simulations of collapsing filaments to form cores, coupled with chemistry, rather than pseudo-time-dependent astrochemical models.
Neither of these studies considered NSC on grains.
The main purpose of this section is not to estimate the most accurate timescale, but to demonstrate that even with a simple model setup similar to \citet{brunken14}, the inclusion of NSC on grains can alter the conclusion.

\begin{figure*}[ht!]
\plotone{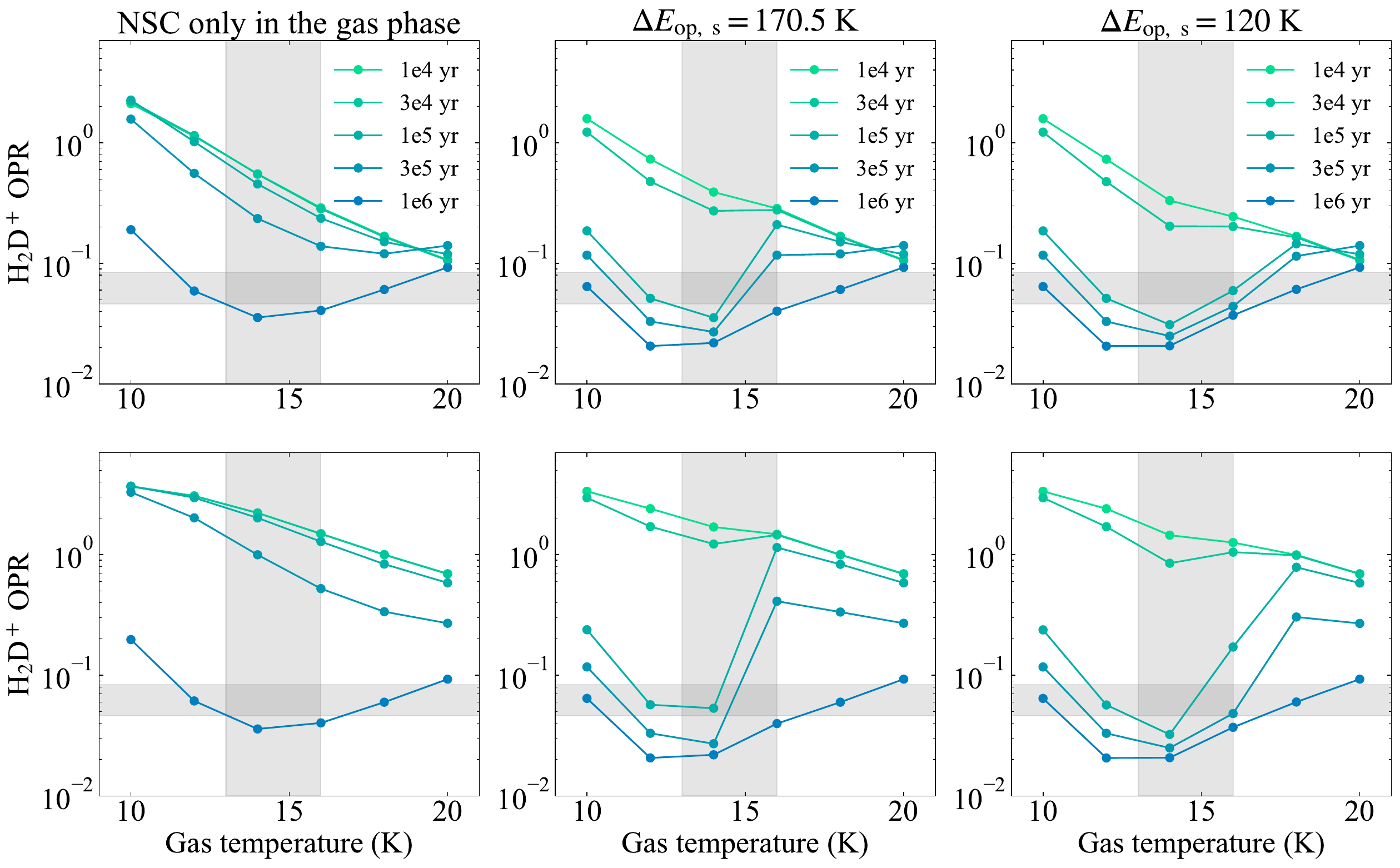}
\caption{Temporal evolution of the \ce{H2D+} OPR predicted in the model without (left) and with the NSC on grains assuming $\Delta E_{\rm op,\,s} = 170.5$ K (middle) or 120 K (right). The gas density is set to 10$^5$ cm$^{-3}$, while the temperature varies from 10 K to 20 K.
The gray areas indicate the \ce{H2D+} OPR and the gas temperature constrained by the observations toward IRAS 16293-2422 \citep{brunken14}.
The upper and lower panels correspond to the cases when the initial \ce{H2} OPR is $10^{-3}$ and 10$^{-2}$, respectively.}
\label{fig:brunken}
\end{figure*}

To check the impact of the NSC on grains on the estimated timescale, we run our astrochemical models described in Section \ref{sec:model} with a similar setting to the model of \citet{brunken14}, but considering the NSC on grains.
We consider three cases with different $\deops$; 170.5 K, 120 K, and 85 K.
In all simulations, the gas density is set to 10$^5$ cm$^{-3}$ and $A_V$ is set to 10 mag.
Gas and dust temperatures are assumed to be the same and vary between 10 K to 20 K.
The initial \ce{H2} OPR is set to 10$^{-3}$ as in \citet{brunken14}.
The upper panels of Figure \ref{fig:brunken} shows the \ce{H2D+} OPR as a function of the temperature at different times.
The simulation results with $\deops = 85$ K are not shown in Fig. \ref{fig:brunken}, as they are essentially similar to those with $\deops = 120$ K.
As found in \citet{brunken14}, when only the NSC in the gas-phase is considered, it takes $\sim$1 Myr to reach the observed level of the \ce{H2D+} OPR.
On the other hand, when the NSC on grains is additionally considered, 0.1 Myr is long enough to explain the observed \ce{H2D+} OPR at the temperature of 12 K - 16 K.
Exceptionally, when $\deops = 170.5$ K and $T=16$ K, NSC on grains is less efficient than the NSC in the gas-phase, and still 1 Myr timescale is required.
The simulation results with $\deops = 85$ K are not shown in Fig. \ref{fig:brunken}, as they are essentially similar to those with $\deops = 120$ K. 
The above results hold even when we assume the initial \ce{H2} OPR of 10$^{-2}$ as shown in the lower panels of the figure.

%Deuterium fractionation ratio, such as \ce{N2D+}/\ce{N2H+}, has often been used as an indirect tracer of the \ce{H2} OPR, because it is difficult to directly observe \ce{H2} in the cold gas ($\sim$10 K),deuterium fractionation is not sensitive to \ce{H2} OPR below 10$^{-3}$.
%pagani results would not be changed even if we consider the NSC on grains.

%\subsection{Caveat}
A caveat for the above discussion is that our model used the experimentally derived binding energy distribution of \ce{H2} and the NSC rate on water ice surfaces.
In star- and planet-forming regions, gaseous \ce{H2} would interact with not only water ice surfaces, but also various types of surfaces, including CO, \ce{CO2}, and \ce{CH3OH} ices.
To the best of our knowledge, similar experimental measurements that are adequate for ices other than water are not available in the literature.
Once such measurements become available, it is straightforward to run similar simulations to those done here.
On the other hand, \citet{kouchi21} has proposed that in molecular clouds, \ce{CO2} nano-crystals are embedded in the amorphous \ce{H2O} ice, and a polyhedral CO crystal is attached to the amorphous \ce{H2O}.
If this is the case, most \ce{H2} could interact with amorphous \ce{H2O} ice even if other species than water exist on grain surfaces.
%The detection of water vapor at the center of L1544 prestellar core supports this scenario \citep{caselli12}.

\subsection{Conditions under which NSC on grains and $\deops$ are important} \label{sec:condition}
As shown above, the NSC on grains is more efficient than that in the gas phase at densities of $\gtrsim$10$^4$ cm$^{-3}$ and temperatures of $\lesssim$14-16 K, depending on $\deops$.
The evolution of the \ce{H2} OPR in lower density regions ($\lesssim$10$^4$ cm$^{-3}$) was studied by \citet{furuya15}, using a simple 1D model of the molecular cloud formation driven by shock compression of atomic gas, and gas-ice chemical network calculations in post-processing.
They found that during the transition from atomic to molecular hydrogen, the NSC in the gas phase is efficient ($\tau_{\rm op,\,g} < \tau_{\rm \ce{H2}}$) and that the \ce{H2} OPR becomes lower than the statistical value of three ($\lesssim$0.1) before the conversion of atomic H into \ce{H2} is nearly complete (see their Appendix C for details).
\citet{lupi21} performed 3D hydrodynamical simulations of collapsing filaments coupled with chemistry and found that the \ce{H2} OPR decrease to $\sim$10$^{-3}$-10$^{-2}$ due to the NSC in the gas phase before the gas density reaches $\sim$10$^4$ cm$^{-3}$.

It should be noted that the evolution of the \ce{H2} OPR via the NSC in the gas phase may be affected by the elemental abundance of gas-phase sulfur \citep{furuya15}.
A higher sulfur abundance reduce the efficiency of the NSC in the gas phase, because \ce{H+}, which is the key species for the NSC, is destroyed by abundant S atoms and CS.
In a specific model of \citet{furuya15}, if the non-depleted (i.e., solar) sulfur elemental abundance is adopted, the \ce{H2} OPR remains high ($>$10$^{-2}$) at the end of their simulations ($\sim$10 Myr after the shock compression), 
whereas adopting a low sulfur elemental abundance (depletion factor of $\sim$100) yields an the \ce{H2} OPR as low as 10$^{-3}$.
The chemical network of \citet{lupi21} did not include any sulfur-bearing species. 
Recent molecular line observations toward molecular clouds indicate that the level of gas-phase sulfur depletion is smaller than previously thought and vary among clouds from a factor of $\sim$20 (Taurus and Perseus molecular clouds) to no depletion (Orion) \citep{fuente23}.
It would be important to simulate the \ce{H2} OPR evolution together with sulfur chemistry using 3D hydrodynamical simulations for better quantifying the \ce{H2} OPR in relatively low density regions of $\lesssim$10$^{4}$ cm$^{-3}$.
%To the best of our knowledge, the effect of sulfur on the \ce{H2} OPR has not explored by 
%Taken together, we think that whether the \ce{H2} OPR can decrease to values as low as $\sim$10$^{-3}$-10$^{-2}$ in molecular clouds remains a matter of debate.

When the \ce{H2} OPR is $\sim$10$^{-3}$, the destruction rate of \ce{H2D+} by \ce{H2} is comparable to that by CO, assuming a CO abundance of $\sim$10$^{-4}$ relative to \ce{H2} \citep{furuya15}. 
Thus,  even if a relatively low \ce{H2} OPR ($\sim$10$^{-2}$ or lower) is established during the formation and evolution of molecular clouds with densities lower than $\sim$10$^4$ cm$^{-3}$, it would still not low enough for extreme deuterium fractionation observed in denser, prestellar cores: the very high D/H ratio of $>$0.1 for e.g., \ce{N2H+} \citep{crapsi05} and \ce{CH3OH} \citep{spezzano25} and abundant doubly deuterated species \citep{Chacon-Tanarro19}.
Further reduction of the \ce{H2} OPR would be required and it would be driven mainly by the NSC on grains (see Section \ref{sec:model}).
%Given that the typical gas density in prestellar cores exceeds 10$^4$ cm$^{-3}$, we argue that NSC on grain surfaces is more important than the gas-phase NSC in this regime. 
An accurate treatment of grain-surface NSC is therefore important for modeling deuterium chemistry in prestellar cores and other dense regions, such as protoplanetary disks \citep{aikawa18}.

On the other hand, in the central regions of prestellar cores without nearby star formation activity, dust temperatures are expected to be below 10 K \citep[e.g.,][]{keto10,hocuk17}. 
At such low temperatures, the exact value of $\deops$ is not important for the efficiency of the NSC.
The $\deops$ value becomes important only in slightly warmer regions ($\sim$15 K), such as the outer envelopes of low-mass protostars (as discussed in Section \ref{sec:discuss}), prestellar cores exposed to strong external radiation fields (e.g., from nearby stars), or the midplane of the outer regions of protoplanetary disks.

%In this work, the temperature is assumed to be constant or determined by thermodynamic equilibrium. 
%Fluctuation of dust temperature by the stochastic heating by UV photons and cosmic rays may be important, because dust temperature at the time when the NSC occurs and that when thermal desorption occurs can be different. 

\section{Summary} \label{sec:summary}
We have investigated how the \ce{H2} OPR and deuterium fractionation in star-forming regions is affected by the NSC on dust grains, focusing on the rotational energy difference between o-\ce{H2} and p-\ce{H2} on grain surfaces ($\deops$).
The previous work by \citet{furuya19} developed a rigorous formulation for the conversion rate of gas-phase o-\ce{H2} (p-\ce{H2}) to gas-phase p-\ce{H2} (o-\ce{H2}) via the NSC on grains, assuming that adsorbed o-\ce{H2} has higher rotational energy than adsorbed p-\ce{H2} by 170.5 K, as in the gas phase.
However, their energy difference can be lower than 170.5 K, because interactions between the surface and adsorbed \ce{H2} molecules affect their rotational motion.
In this work, we relax the assumption in \citet{furuya19} and re-evaluate the conversion rate between o-\ce{H2} and p-\ce{H2}.
We have found that the conversion rate depends on $\deops$. 
On the other hand the steady-state \ce{H2} OPR determined solely by the NSC on grains is largely insensitive to $\deops$, because two effects ($\deops \leq \deopg$ and higher binding energy of o-\ce{H2} than p-\ce{H2}) are canceled out, even when various different binding sites and thermal hopping among them are considered.
Therefore, regardless of $\deops$, one can include the NSC on grains in any astrochemical kinetic models with the slightly modified version of the formulation by \citet{furuya19} (Eqs. \ref{eq:op_surf1}, \ref{eq:op_surf2}, and \ref{eq:op_surf4}).
%while the yield of gaseous p-\ce{H2} per o-\ce{H2} adsorption (the parameter $\eta_{\rm op}$ in Eq.\ref{eq:op_surf1}) depends on $\deops$.
%If $\deops = 85$ K, the conversion efficiency by the NSC on grains is more efficient than that by the NSC via the gas phase reactions even at 16 K.
When the dust temperature is $\lesssim$14-16 K (depending on $\deops$), and the gas density is higher than 10$^4$ cm$^{-3}$, the NSC on grains can be more important than that via the gas-phase proton-exchange reactions (see Sections \ref{sec:steady_state} and \ref{sec:condition} for details).

Finally, we have added the NSC on grain surfaces to the existing gas-ice astrochemical model with deuterated species and the nuclear spin states of light species.
As case studies, we have investigated the time evolution of the \ce{H2} OPR and deuterium fractionation in collapsing prestellar cores and in the outer envelope of the low mass protostar.
In the prestellar core model, the deuterium fractionation is more efficient at $\gtrsim$10$^5$ cm$^{-3}$ in the case when the NSC on grains compared to the case without the NSC on grains (the D/H ratio for \ce{H3+} isotopologues can be enhanced by a factor of 10 or more).
In addition, the dependence of the D/H ratio for \ce{H3+} isotopologues on the inital \ce{H2} OPR becomes weaker by considering the NSC on grains. 
In the protostellar envelope model, we have found that by considering NSC on grains, 10$^5$ yr is long enough for the \ce{H2D+} OPR to reach the observed value by \citet{brunken14}, while when only NSC in the gas phase is considered, $\sim$10$^6$ yr is required to explain the observed level of the \ce{H2D+} OPR as claimed by \citet{brunken14}.
%Then, considering the NSC on grains is important for accurately modeling the deuterium fractionation chemistry in dense, cold regions of the ISM.
%, and it also reduces the uncertainty of the deuterium fractionation due to the uncertainty of the ``initial'' \ce{H2} OPR.
%To use the \ce{H2} OPR or \ce{H2D+} OPR as a chemical clock, the effect of the NSC on grains should be taken into account.

\acknowledgments
We would like to thank the anonymous referees for their useful comments.
This work is supported in part by JSPS KAKENHI Grant numbers 20H05847, 22H00296, 24K00686, and 25K07364.

\software{Matplotlib \citep{matplotlib}}

%\begin{thebibliography}
%\end{thebibliography}



\begin{appendix}
%\section{Steady state \ce{H2D+}/\ce{H3+}, \ce{D2H+}/\ce{H2D+}, and \ce{D3+}/\ce{D2H+} ratios}
\section{Energy difference between adsorbed ortho and para hydrogen}
\label{app:energy_difference}
The surface potential $V(z, \theta)$ of a molecule is mainly characterized by its well-depth and anisotropy as functions of the distance ($z$) between the mass center of the molecule and the surface, and the molecular-axis angle ($\theta$) with respect to the surface normal. When the degree of anisotropy is not so large, the surface potential is simply approximated as $V(z,\theta)=V_0(z)+V_2P_2(\cos \theta)$, where $P_2(\cos \theta)$ is the second order Legendre polynomial, and $V_0(z)$ and $V_2$ are the isotropic surface potential and the degree of anisotropy, respectively \citep{yu85,whaley85}.
Moreover, when the anisotropic surface field is not so large, the $V_2P_2(\cos \theta)$ term can be treated as the first-order perturbation relative to the free molecular rotation in the isotropic $V_0(z)$ potential. Within the first-order perturbation, the wave function is regarded unaltered, whereas the rotational energy shift is represented as follows \citep{andersson82,nordlander85}:
\begin{linenomath*}
\begin{align}
\Delta E_{\rm shift}(J,m) &= V_2 \braket{J, m|P_2(\cos \theta)|J,m} \nonumber \\
&= \frac{3}{2J+3}\left(\frac{J^2-m^2}{2J-1} - \frac{J}{3} \right)V_2,
\end{align}
\end{linenomath*}
where $\ket{J,m}$ is the rotational wave function written by spherical harmonics $Y_{J,m}(\theta, \phi)$ and $m$ is the $z$ component of $J$.
The rotational matrix element is given in \citet{andersson82,nordlander85,fukutani13}.
Whereas the $J=0$ state (p-\ce{H2}) shows no shift, the $m$ degeneracy of $J=1$ state (o-\ce{H2}) is lifted with an energy splitting of $3V_2/5$; the $(J,m)=(1,0)$ and (1,$\pm$1) rotational states are shifted by $\Delta E_{\rm shift}(1, 0) = 2V_2/5$ and $\Delta E_{\rm shift}(1, \pm1) = -V_2/5$, respectively (Figure \ref{fig:erot}). 
When $V_2$ is negative, the ground state of o-\ce{H2} is $(J,m)=(1,0)$ (i.e., no rotational degeneracy), and the energy difference between the ground states of p-\ce{H2} and o-\ce{H2} is $\Delta E_{\rm op,\,s}= \deopg + \Delta E_{\rm shift}(1, 0) = 2B+2V_2/5$ (Figure \ref{fig:erot}).
In contrast, when $V_2$ is positive, the ground state of o-\ce{H2} is $(J,m)=(1,\pm1)$ (the rotational degeneracy is 2), and the energy difference between the ground states of p-\ce{H2} and o-\ce{H2} is $\Delta E_{\rm op,\,s} = 2B-V_2/5$ (Figure \ref{fig:erot}).
Depending on the sign of $V_2$, the rotational motion tends to be restricted to either the surface-normal direction (1D hindered rotation) for negative $V_2$, or the surface-parallel direction (2D hindered rotation) for positive $V_2$.

In the case of metal surfaces, e.g., Ag(111) and Au(110), $V_2$ is around -5 meV \citep{sugimoto14,sugimoto17}.
Then the rotational energy difference between the lowest rotational level of p-\ce{H2} and o-\ce{H2} is estimated to be $\Delta E_{\rm op,\,s} \sim 12.8$ meV ($\sim$150 K) that is larger than $\Delta E_{\rm op,\,2Drot} \sim 7.4$ meV ($\sim$85 K) but smaller than $\deopg \sim 14.8$ meV ($\sim$170 K).
As ASW is more highly anisotropic than metals, $\Delta E_{\rm op,\,s}$ on ASW would be somewhere between that on the metals and $\Delta E_{\rm op,\,2Drot}$.

\end{appendix}

\bibliography{ref_h2opr}{}

@ARTICLE{bron16,
       author = {{Bron}, Emeric and {Le Petit}, Franck and {Le Bourlot}, Jacques},
        title = "{Efficient ortho-para conversion of H$_{2}$ on interstellar grain surfaces}",
      journal = {\aap},
         year = 2016,
        month = apr,
       volume = {588},
          eid = {A27},
        pages = {A27},
          doi = {10.1051/0004-6361/201527879},
archivePrefix = {arXiv},
       eprint = {1601.04356},
 primaryClass = {astro-ph.GA},
       adsurl = {https://ui.adsabs.harvard.edu/abs/2016A&A...588A..27B}
}

@ARTICLE{spezzano25,
       author = {{Spezzano}, S. and {Riedel}, W. and {Caselli}, P. and {Sipil{\"a}}, O. and {Lin}, Y. and {Bunn}, H.~A. and {Redaelli}, E. and {Coudert}, L.~H. and {Meg{\'\i}as}, A. and {Jimenez-Serra}, I.},
        title = "{High deuteration of methanol in L1544}",
      journal = {arXiv e-prints},
         year = 2025,
        month = dec,
          eid = {arXiv:2512.09110},
        pages = {arXiv:2512.09110},
          doi = {10.48550/arXiv.2512.09110},
archivePrefix = {arXiv},
       eprint = {2512.09110},
 primaryClass = {astro-ph.GA},
       adsurl = {https://ui.adsabs.harvard.edu/abs/2025arXiv251209110S}
}

@ARTICLE{gerlich90,
       author = {{Gerlich}, D.},
        title = "{Ortho-para transitions in reactive H$^{ + }$+H$_{2}$ collisions}",
      journal = {\jcp},
         year = 1990,
        month = feb,
       volume = {92},
       number = {4},
        pages = {2377-2388},
          doi = {10.1063/1.457980},
       adsurl = {https://ui.adsabs.harvard.edu/abs/1990JChPh..92.2377G}
}

@ARTICLE{le_Bourlot91,
       author = {{Le Bourlot}, J.},
        title = "{Ammonia formation and the ortho-to-para ratio of H2 in dark clouds.}",
      journal = {\aap},
         year = 1991,
        month = feb,
       volume = {242},
        pages = {235},
       adsurl = {https://ui.adsabs.harvard.edu/abs/1991A&A...242..235L}
}

@INPROCEEDINGS{dunham14,
       author = {{Dunham}, M.~M. and {Stutz}, A.~M. and {Allen}, L.~E. and {Evans}, II, N.~J. and {Fischer}, W.~J. and {Megeath}, S.~T. and {Myers}, P.~C. and {Offner}, S.~S.~R. and {Poteet}, C.~A. and {Tobin}, J.~J. and {Vorobyov}, E.~I.},
        title = "{The Evolution of Protostars: Insights from Ten Years of Infrared Surveys with Spitzer and Herschel}",
    booktitle = {Protostars and Planets VI},
         year = 2014,
       editor = {{Beuther}, Henrik and {Klessen}, Ralf S. and {Dullemond}, Cornelis P. and {Henning}, Thomas},
        month = jan,
        pages = {195-218},
          doi = {10.2458/azu_uapress_9780816531240-ch009},
archivePrefix = {arXiv},
       eprint = {1401.1809},
 primaryClass = {astro-ph.GA},
       adsurl = {https://ui.adsabs.harvard.edu/abs/2014prpl.conf..195D}
}

@BOOK{spitzer78,
       author = {{Spitzer}, Lyman},
        title = "{Physical processes in the interstellar medium}",
         year = 1978,
          doi = {10.1002/9783527617722},
       adsurl = {https://ui.adsabs.harvard.edu/abs/1978ppim.book.....S}
}

@ARTICLE{furuya24,
       author = {{Furuya}, Kenji},
        title = "{A Framework for Incorporating Binding Energy Distribution in Gas-ice Astrochemical Models}",
      journal = {\apj},
         year = 2024,
        month = oct,
       volume = {974},
       number = {1},
          eid = {115},
        pages = {115},
          doi = {10.3847/1538-4357/ad6b20},
archivePrefix = {arXiv},
       eprint = {2408.02958},
 primaryClass = {astro-ph.GA},
       adsurl = {https://ui.adsabs.harvard.edu/abs/2024ApJ...974..115F}
}

@ARTICLE{Jimenez-Redondo24,
       author = {{Jim{\'e}nez-Redondo}, Miguel and {Sipil{\"a}}, Olli and {Jusko}, Pavol and {Caselli}, Paola},
        title = "{Measurements and simulations of rate coefficients for the deuterated forms of the H$_{2}$$^{+}$ + H$_{2}$ and H$_{3}$$^{+}$ + H$_{2}$ reactive systems at low temperature}",
      journal = {\aap},
         year = 2024,
        month = dec,
       volume = {692},
          eid = {A121},
        pages = {A121},
          doi = {10.1051/0004-6361/202451757},
archivePrefix = {arXiv},
       eprint = {2412.02206},
 primaryClass = {astro-ph.SR},
       adsurl = {https://ui.adsabs.harvard.edu/abs/2024A&A...692A.121J}
}

@ARTICLE{he14,
       author = {{He}, Jiao and {Vidali}, Gianfranco},
        title = "{Application of a diffusion-desorption rate equation model in astrochemistry}",
      journal = {Faraday Discussions},
         year = 2014,
        month = jan,
       volume = {168},
        pages = {517},
          doi = {10.1039/C3FD00113J},
       adsurl = {https://ui.adsabs.harvard.edu/abs/2014FaDi..168..517H}
}

@article{yu85,
    author = {Yu, Chien‐fan and Whaley, K. Birgitta and Hogg, C. S. and Sibener, S. J.},
    title = "{Investigation of the spatially isotropic component of the laterally averaged molecular hydrogen/Ag(111) physisorption potential}",
    journal = {The Journal of Chemical Physics},
    volume = {83},
    number = {8},
    pages = {4217-4234},
    year = {1985},
    month = {10},
    issn = {0021-9606},
    doi = {10.1063/1.449086},
    url = {https://doi.org/10.1063/1.449086},
    eprint = {https://pubs.aip.org/aip/jcp/article-pdf/83/8/4217/10986234/4217\_1\_online.pdf},
}

@article{whaley85,
    author = {Whaley, K. Birgitta and Yu, Chien‐fan and Hogg, C. S. and Light, John C. and Sibener, S. J.},
    title = "{Investigation of the spatially anisotropic component of the laterally averaged molecular hydrogen/Ag(111) physisorption potential}",
    journal = {The Journal of Chemical Physics},
    volume = {83},
    number = {8},
    pages = {4235-4255},
    year = {1985},
    month = {10},
    issn = {0021-9606},
    doi = {10.1063/1.449087},
    url = {https://doi.org/10.1063/1.449087},
    eprint = {https://pubs.aip.org/aip/jcp/article-pdf/83/8/4235/10987286/4235\_1\_online.pdf},
}

@article{nordlander85,
title = {Physisorption interaction of H2 with noble metals},
journal = {Surface Science},
volume = {152-153},
pages = {702-709},
year = {1985},
issn = {0039-6028},
doi = {https://doi.org/10.1016/0039-6028(85)90478-9},
url = {https://www.sciencedirect.com/science/article/pii/0039602885904789},
author = {Peter Nordlander and Claes Holmberg and John Harris}
}

@article{andersson82,
  title = {Observation of Rotational Transitions for ${\mathrm{H}}_{2}$, ${\mathrm{D}}_{2}$, and HD Adsorbed on Cu(100)},
  author = {Andersson, S. and Harris, J.},
  journal = {Phys. Rev. Lett.},
  volume = {48},
  issue = {8},
  pages = {545--548},
  numpages = {0},
  year = {1982},
  month = {Feb},
  publisher = {American Physical Society},
  doi = {10.1103/PhysRevLett.48.545},
  url = {https://link.aps.org/doi/10.1103/PhysRevLett.48.545}
}

@article{sugimoto14,
  title = {Effects of Rotational-Symmetry Breaking on Physisorption of Ortho- and Para-${\mathrm{H}}_{2}$ on Ag(111)},
  author = {Sugimoto, Toshiki and Fukutani, Katsuyuki},
  journal = {Phys. Rev. Lett.},
  volume = {112},
  issue = {14},
  pages = {146101},
  numpages = {5},
  year = {2014},
  month = {Apr},
  publisher = {American Physical Society},
  doi = {10.1103/PhysRevLett.112.146101},
  url = {https://link.aps.org/doi/10.1103/PhysRevLett.112.146101}
}

@article{sugimoto17,
  title = {Inelastic electron tunneling mediated by a molecular quantum rotator},
  author = {Sugimoto, Toshiki and Kunisada, Yuji and Fukutani, Katsuyuki},
  journal = {Phys. Rev. B},
  volume = {96},
  issue = {24},
  pages = {241409},
  numpages = {7},
  year = {2017},
  month = {Dec},
  publisher = {American Physical Society},
  doi = {10.1103/PhysRevB.96.241409},
  url = {https://link.aps.org/doi/10.1103/PhysRevB.96.241409}
}

@ARTICLE{fukutani13,
       author = {{Fukutani}, K. and {Sugimoto}, T.},
        title = "{Physisorption and ortho-para conversion of molecular hydrogen on solid surfaces}",
      journal = {Progress In Surface Science},
         year = 2013,
        month = dec,
       volume = {88},
       number = {4},
        pages = {279-348},
          doi = {10.1016/j.progsurf.2013.09.001},
       adsurl = {https://ui.adsabs.harvard.edu/abs/2013PrSS...88..279F}
}

@ARTICLE{pagani92,
       author = {{Pagani}, L. and {Salez}, M. and {Wannier}, P.~G.},
        title = "{The chemistry of H2D+ in cold clouds.}",
      journal = {\aap},
         year = 1992,
        month = may,
       volume = {258},
        pages = {479-488},
       adsurl = {https://ui.adsabs.harvard.edu/abs/1992A&A...258..479P}
}

@ARTICLE{maret07,
       author = {{Maret}, S{\'e}bastien and {Bergin}, Edwin A.},
        title = "{The Ionization Fraction of Barnard 68: Implications for Star and Planet Formation}",
      journal = {\apj},
         year = 2007,
        month = aug,
       volume = {664},
       number = {2},
        pages = {956-963},
          doi = {10.1086/519152},
archivePrefix = {arXiv},
       eprint = {0704.3188},
 primaryClass = {astro-ph},
       adsurl = {https://ui.adsabs.harvard.edu/abs/2007ApJ...664..956M}
}

@ARTICLE{tsuge19,
       author = {{Tsuge}, Masashi and {Hama}, Tetsuya and {Kimura}, Yuki and {Kouchi}, Akira and {Watanabe}, Naoki},
        title = "{Interactions of Atomic and Molecular Hydrogen with a Diamond-like Carbon Surface: H$_{2}$ Formation and Desorption}",
      journal = {\apj},
         year = 2019,
        month = jun,
       volume = {878},
       number = {1},
          eid = {23},
        pages = {23},
          doi = {10.3847/1538-4357/ab1e4e},
archivePrefix = {arXiv},
       eprint = {1911.05300},
 primaryClass = {astro-ph.GA},
       adsurl = {https://ui.adsabs.harvard.edu/abs/2019ApJ...878...23T}
}

@ARTICLE{ueta16,
       author = {{Ueta}, Hirokazu and {Watanabe}, Naoki and {Hama}, Tetsuya and {Kouchi}, Akira},
        title = "{Surface Temperature Dependence of Hydrogen Ortho-Para Conversion on Amorphous Solid Water}",
      journal = {\prl},
         year = 2016,
        month = jun,
       volume = {116},
       number = {25},
          eid = {253201},
        pages = {253201},
          doi = {10.1103/PhysRevLett.116.253201},
       adsurl = {https://ui.adsabs.harvard.edu/abs/2016PhRvL.116y3201U}
}

@ARTICLE{tsuge21a,
       author = {{Tsuge}, Masashi and {Kouchi}, Akira and {Watanabe}, Naoki},
        title = "{Measurements of Ortho-to-para Nuclear Spin Conversion of H$_{2}$ on Low-temperature Carbonaceous Grain Analogs: Diamond-like Carbon and Graphite}",
      journal = {\apj},
         year = 2021,
        month = dec,
       volume = {923},
       number = {1},
          eid = {71},
        pages = {71},
          doi = {10.3847/1538-4357/ac2a33},
archivePrefix = {arXiv},
       eprint = {2109.12734},
 primaryClass = {astro-ph.GA},
       adsurl = {https://ui.adsabs.harvard.edu/abs/2021ApJ...923...71T}
}

@ARTICLE{bovino17,
       author = {{Bovino}, S. and {Grassi}, T. and {Schleicher}, D.~R.~G. and {Caselli}, P.},
        title = "{H$_{2}$ Ortho-to-para Conversion on Grains: A Route to Fast Deuterium Fractionation in Dense Cloud Cores?}",
      journal = {\apjl},
         year = 2017,
        month = nov,
       volume = {849},
       number = {2},
          eid = {L25},
        pages = {L25},
          doi = {10.3847/2041-8213/aa95b7},
archivePrefix = {arXiv},
       eprint = {1708.02046},
 primaryClass = {astro-ph.GA},
       adsurl = {https://ui.adsabs.harvard.edu/abs/2017ApJ...849L..25B}
}

@ARTICLE{hilyblant18,
       author = {{Hily-Blant}, P. and {Faure}, A. and {Rist}, C. and {Pineau des For{\^e}ts}, G. and {Flower}, D.~R.},
        title = "{Modelling the molecular composition and nuclear-spin chemistryof collapsing pre-stellar sources}",
      journal = {\mnras},
         year = 2018,
        month = jul,
       volume = {477},
       number = {4},
        pages = {4454-4472},
          doi = {10.1093/mnras/sty881},
archivePrefix = {arXiv},
       eprint = {1804.01354},
 primaryClass = {astro-ph.GA},
       adsurl = {https://ui.adsabs.harvard.edu/abs/2018MNRAS.477.4454H}
}

@ARTICLE{flower06,
       author = {{Flower}, D.~R. and {Pineau Des For{\^e}ts}, G. and {Walmsley}, C.~M.},
        title = "{The importance of the ortho:para H$_{2}$ ratio for the deuteration of molecules during pre-protostellar collapse}",
      journal = {\aap},
         year = 2006,
        month = apr,
       volume = {449},
       number = {2},
        pages = {621-629},
          doi = {10.1051/0004-6361:20054246},
archivePrefix = {arXiv},
       eprint = {astro-ph/0601429},
 primaryClass = {astro-ph},
       adsurl = {https://ui.adsabs.harvard.edu/abs/2006A&A...449..621F}
}

@ARTICLE{sipila13,
       author = {{Sipil{\"a}}, O. and {Caselli}, P. and {Harju}, J.},
        title = "{HD depletion in starless cores}",
      journal = {\aap},
         year = 2013,
        month = jun,
       volume = {554},
          eid = {A92},
        pages = {A92},
          doi = {10.1051/0004-6361/201220922},
archivePrefix = {arXiv},
       eprint = {1304.4031},
 primaryClass = {astro-ph.GA},
       adsurl = {https://ui.adsabs.harvard.edu/abs/2013A&A...554A..92S}
}

@ARTICLE{tsuge21b,
       author = {{Tsuge}, M. and {Namiyoshi}, T. and {Furuya}, K. and {Yamazaki}, T. and {Kouchi}, A. and {Watanabe}, N.},
        title = "{Rapid Ortho-to-para Nuclear Spin Conversion of H$_{2}$ on a Silicate Dust Surface}",
      journal = {\apj},
         year = 2021,
        month = feb,
       volume = {908},
       number = {2},
          eid = {234},
        pages = {234},
          doi = {10.3847/1538-4357/abd9c0},
archivePrefix = {arXiv},
       eprint = {2101.02357},
 primaryClass = {astro-ph.GA},
       adsurl = {https://ui.adsabs.harvard.edu/abs/2021ApJ...908..234T}
}

@ARTICLE{pagani11,
       author = {{Pagani}, Laurent and {Roueff}, Evelyne and {Lesaffre}, Pierre},
        title = "{Ortho-H$_{2}$ and the Age of Interstellar Dark Clouds}",
      journal = {\apjl},
         year = 2011,
        month = oct,
       volume = {739},
       number = {2},
          eid = {L35},
        pages = {L35},
          doi = {10.1088/2041-8205/739/2/L35},
archivePrefix = {arXiv},
       eprint = {1109.6495},
 primaryClass = {astro-ph.GA},
       adsurl = {https://ui.adsabs.harvard.edu/abs/2011ApJ...739L..35P}
}

@ARTICLE{cooper19,
       author = {{Cooper}, April M. and {K{\"a}stner}, Johannes},
        title = "{Low-Temperature Kinetic Isotope Effects in CH3OH + H {\textrightarrow} CH2OH + H2 Shed Light on the Deuteration of Methanol in Space}",
      journal = {Journal of Physical Chemistry A},
         year = 2019,
        month = oct,
       volume = {123},
       number = {42},
        pages = {9061-9068},
          doi = {10.1021/acs.jpca.9b07013},
archivePrefix = {arXiv},
       eprint = {2009.04308},
 primaryClass = {physics.chem-ph},
       adsurl = {https://ui.adsabs.harvard.edu/abs/2019JPCA..123.9061C}
}

@ARTICLE{harju17,
       author = {{Harju}, Jorma and {Sipil{\"a}}, Olli and {Br{\"u}nken}, Sandra and {Schlemmer}, Stephan and {Caselli}, Paola and {Juvela}, Mika and {Menten}, Karl M. and {Stutzki}, J{\"u}rgen and {Asvany}, Oskar and {Kami{\'n}ski}, Tomasz and {Okada}, Yoko and {Higgins}, Ronan},
        title = "{Detection of Interstellar Ortho-D$_{2}$H$^{+}$ with SOFIA}",
      journal = {\apj},
         year = 2017,
        month = may,
       volume = {840},
       number = {2},
          eid = {63},
        pages = {63},
          doi = {10.3847/1538-4357/aa6c69},
archivePrefix = {arXiv},
       eprint = {1704.02526},
 primaryClass = {astro-ph.GA},
       adsurl = {https://ui.adsabs.harvard.edu/abs/2017ApJ...840...63H}
}

@ARTICLE{pagani13,
       author = {{Pagani}, L. and {Lesaffre}, P. and {Jorfi}, M. and {Honvault}, P. and {Gonz{\'a}lez-Lezana}, T. and {Faure}, A.},
        title = "{Ortho-H$_{2}$ and the age of prestellar cores}",
      journal = {\aap},
         year = 2013,
        month = mar,
       volume = {551},
          eid = {A38},
        pages = {A38},
          doi = {10.1051/0004-6361/201117161},
       adsurl = {https://ui.adsabs.harvard.edu/abs/2013A&A...551A..38P}
}

@ARTICLE{brunken14,
       author = {{Br{\"u}nken}, Sandra and {Sipil{\"a}}, Olli and {Chambers}, Edward T. and {Harju}, Jorma and {Caselli}, Paola and {Asvany}, Oskar and {Honingh}, Cornelia E. and {Kami{\'n}ski}, Tomasz and {Menten}, Karl M. and {Stutzki}, J{\"u}rgen and {Schlemmer}, Stephan},
        title = "{H$_{2}$D$^{+}$ observations give an age of at least one million years for a cloud core forming Sun-like stars}",
      journal = {\nat},
         year = 2014,
        month = dec,
       volume = {516},
       number = {7530},
        pages = {219-221},
          doi = {10.1038/nature13924},
       adsurl = {https://ui.adsabs.harvard.edu/abs/2014Natur.516..219B}
}

@ARTICLE{linsky03,
       author = {{Linsky}, Jeffrey L.},
        title = "{Atomic Deuterium/Hydrogen in the Galaxy}",
      journal = {\ssr},
         year = 2003,
        month = apr,
       volume = {106},
       number = {1},
        pages = {49-60},
          doi = {10.1023/A:1024673217736},
archivePrefix = {arXiv},
       eprint = {astro-ph/0309099},
 primaryClass = {astro-ph},
       adsurl = {https://ui.adsabs.harvard.edu/abs/2003SSRv..106...49L}
}

@ARTICLE{honvault12,
       author = {{Honvault}, P. and {Jorfi}, M. and {Gonz{\'a}lez-Lezana}, T. and {Faure}, A. and {Pagani}, L.},
        title = "{Erratum: Otho-Para H$_{2}$ Conversion by Proton Exchange at Low Temperature: An Accurate Quantum Mechanical Study [Phys. Rev. Lett. 107, 023201 (2011)]}",
      journal = {\prl},
         year = 2012,
        month = mar,
       volume = {108},
       number = {10},
          eid = {109903},
        pages = {109903},
          doi = {10.1103/PhysRevLett.108.109903},
       adsurl = {https://ui.adsabs.harvard.edu/abs/2012PhRvL.108j9903H}
}

@ARTICLE{millar89,
       author = {{Millar}, T.~J. and {Bennett}, A. and {Herbst}, Eric},
        title = "{Deuterium Fractionation in Dense Interstellar Clouds}",
      journal = {\apj},
         year = 1989,
        month = may,
       volume = {340},
        pages = {906},
          doi = {10.1086/167444},
       adsurl = {https://ui.adsabs.harvard.edu/abs/1989ApJ...340..906M}
}

@ARTICLE{watson76,
       author = {{Watson}, William D.},
        title = "{Interstellar molecule reactions}",
      journal = {Reviews of Modern Physics},
         year = 1976,
        month = oct,
       volume = {48},
       number = {4},
        pages = {513-552},
          doi = {10.1103/RevModPhys.48.513},
       adsurl = {https://ui.adsabs.harvard.edu/abs/1976RvMP...48..513W}
}

@ARTICLE{hillenbrand19,
       author = {{Hillenbrand}, P. -M. and {Bowen}, K.~P. and {Li{\'e}vin}, J. and {Urbain}, X. and {Savin}, D.~W.},
        title = "{}",
      journal = {\apj},
         year = 2019,
        month = may,
       volume = {877},
       number = {1},
          eid = {38},
        pages = {38},
          doi = {10.3847/1538-4357/ab16dc},
archivePrefix = {arXiv},
       eprint = {1904.02955},
 primaryClass = {astro-ph.IM},
       adsurl = {https://ui.adsabs.harvard.edu/abs/2019ApJ...877...38H}
}

@article{brown89,
    author = {Brown, Paul D. and Millar, T. J.},
    title = "{Models of the gas?grain interaction ? deuterium chemistry}",
    journal = {Monthly Notices of the Royal Astronomical Society},
    volume = {237},
    number = {3},
    pages = {661-671},
    year = {1989},
    month = {04},
    issn = {0035-8711},
    doi = {10.1093/mnras/237.3.661},
    url = {https://doi.org/10.1093/mnras/237.3.661},
    eprint = {https://academic.oup.com/mnras/article-pdf/237/3/661/3040059/mnras237-0661.pdf},
}

@ARTICLE{gerlich02,
       author = {{Gerlich}, Dieter and {Herbst}, Eric and {Roueff}, Evelyne},
        title = "{H$_{3}$$^{+}$+HD<-- >H$_{2}$D$^{+}$+H$_{2}$: low-temperature laboratory measurements and interstellar implications}",
      journal = {\planss},
         year = 2002,
        month = oct,
       volume = {50},
       number = {12-13},
        pages = {1275-1285},
          doi = {10.1016/S0032-0633(02)00094-6},
       adsurl = {https://ui.adsabs.harvard.edu/abs/2002P&SS...50.1275G}
}

@ARTICLE{honvault11,
       author = {{Honvault}, P. and {Jorfi}, M. and {Gonz{\'a}lez-Lezana}, T. and {Faure}, A. and {Pagani}, L.},
        title = "{Ortho-Para H$_{2}$ Conversion by Proton Exchange at Low Temperature: An Accurate Quantum Mechanical Study}",
      journal = {\prl},
         year = 2011,
        month = jul,
       volume = {107},
       number = {2},
          eid = {023201},
        pages = {023201},
          doi = {10.1103/PhysRevLett.107.023201},
       adsurl = {https://ui.adsabs.harvard.edu/abs/2011PhRvL.107b3201H}
}

@ARTICLE{kouchi21,
       author = {{Kouchi}, Akira and {Tsuge}, Masashi and {Hama}, Tetsuya and {Oba}, Yasuhiro and {Okuzumi}, Satoshi and {Sirono}, Sin-iti and {Momose}, Munetake and {Nakatani}, Naoki and {Furuya}, Kenji and {Shimonishi}, Takashi and {Yamazaki}, Tomoya and {Hidaka}, Hiroshi and {Kimura}, Yuki and {Murata}, Ken-ichiro and {Fujita}, Kazuyuki and {Nakatsubo}, Shunichi and {Tachibana}, Shogo and {Watanabe}, Naoki},
        title = "{Transmission Electron Microscopy Study of the Morphology of Ices Composed of H$_{2}$O, CO$_{2}$, and CO on Refractory Grains}",
      journal = {\apj},
         year = 2021,
        month = sep,
       volume = {918},
       number = {2},
          eid = {45},
        pages = {45},
          doi = {10.3847/1538-4357/ac0ae6},
archivePrefix = {arXiv},
       eprint = {2109.03404},
 primaryClass = {astro-ph.GA},
       adsurl = {https://ui.adsabs.harvard.edu/abs/2021ApJ...918...45K}
}

@ARTICLE{watanabe10,
       author = {{Watanabe}, Naoki and {Kimura}, Yuki and {Kouchi}, Akira and {Chigai}, Takeshi and {Hama}, Tetsuya and {Pirronello}, Valerio},
        title = "{Direct Measurements of Hydrogen Atom Diffusion and the Spin Temperature of Nascent H$_{2}$ Molecule on Amorphous Solid Water}",
      journal = {\apjl},
         year = 2010,
        month = may,
       volume = {714},
       number = {2},
        pages = {L233-L237},
          doi = {10.1088/2041-8205/714/2/L233},
       adsurl = {https://ui.adsabs.harvard.edu/abs/2010ApJ...714L.233W}
}

@ARTICLE{tielens83,
       author = {{Tielens}, A.~G.~G.~M.},
        title = "{Surface chemistry of deuterated molecules}",
      journal = {\aap},
         year = 1983,
        month = mar,
       volume = {119},
       number = {2},
        pages = {177-184},
       adsurl = {https://ui.adsabs.harvard.edu/abs/1983A&A...119..177T}
}

@ARTICLE{roueff13,
       author = {{Roueff}, Evelyne and {Gerin}, Maryvonne and {Lis}, Dariusz C. and {Wootten}, Alwyn and {Marcelino}, Nuria and {Cernicharo}, Jose and {Tercero}, Belen},
        title = "{CH2D+, the Search for the Holy Grail}",
      journal = {Journal of Physical Chemistry A},
         year = 2013,
        month = oct,
       volume = {117},
       number = {39},
        pages = {9959-9967},
          doi = {10.1021/jp400119a},
archivePrefix = {arXiv},
       eprint = {1306.6795},
 primaryClass = {astro-ph.GA},
       adsurl = {https://ui.adsabs.harvard.edu/abs/2013JPCA..117.9959R}
}

@ARTICLE{ratajczak09,
       author = {{Ratajczak}, A. and {Quirico}, E. and {Faure}, A. and {Schmitt}, B. and {Ceccarelli}, C.},
        title = "{Hydrogen/deuterium exchange in interstellar ice analogs}",
      journal = {\aap},
         year = 2009,
        month = mar,
       volume = {496},
       number = {2},
        pages = {L21-L24},
          doi = {10.1051/0004-6361/200911679},
       adsurl = {https://ui.adsabs.harvard.edu/abs/2009A&A...496L..21R}
}

@ARTICLE{hugo09,
       author = {{Hugo}, Edouard and {Asvany}, Oskar and {Schlemmer}, Stephan},
        title = "{H$_{3}$$^{+}$+H$_{2}$ isotopic system at low temperatures: Microcanonical model and experimental study}",
      journal = {\jcp},
         year = 2009,
        month = apr,
       volume = {130},
       number = {16},
        pages = {164302-164302},
          doi = {10.1063/1.3089422},
       adsurl = {https://ui.adsabs.harvard.edu/abs/2009JChPh.130p4302H}
}

@ARTICLE{goldsmith05,
       author = {{Goldsmith}, P.~F. and {Li}, D.},
        title = "{H I Narrow Self-Absorption in Dark Clouds: Correlations with Molecular Gas and Implications for Cloud Evolution and Star Formation}",
      journal = {\apj},
         year = 2005,
        month = apr,
       volume = {622},
       number = {2},
        pages = {938-958},
          doi = {10.1086/428032},
archivePrefix = {arXiv},
       eprint = {astro-ph/0412427},
 primaryClass = {astro-ph},
       adsurl = {https://ui.adsabs.harvard.edu/abs/2005ApJ...622..938G}
}

@BOOK{tielens05,
       author = {{Tielens}, A.~G.~G.~M.},
        title = "{The Physics and Chemistry of the Interstellar Medium}",
         year = 2005,
       adsurl = {https://ui.adsabs.harvard.edu/abs/2005pcim.book.....T}
}

@ARTICLE{sipila17,
       author = {{Sipil{\"a}}, O. and {Harju}, J. and {Caselli}, P.},
        title = "{Species-to-species rate coefficients for the H$_{3}$$^{+}$ + H$_{2}$ reacting system}",
      journal = {\aap},
         year = 2017,
        month = oct,
       volume = {607},
          eid = {A26},
        pages = {A26},
          doi = {10.1051/0004-6361/201731039},
archivePrefix = {arXiv},
       eprint = {1707.03170},
 primaryClass = {astro-ph.GA},
       adsurl = {https://ui.adsabs.harvard.edu/abs/2017A&A...607A..26S}
}

@INPROCEEDINGS{nomura23,
       author = {{Nomura}, H. and {Furuya}, K. and {Cordiner}, M.~A. and {Charnley}, S.~B. and {Alexander}, C.~M. O'D. and {Nixon}, C.~A. and {Guzman}, V.~V. and {Yurimoto}, H. and {Tsukagoshi}, T. and {Iino}, T.},
        title = "{The Isotopic Links from Planet Forming Regions to the Solar System}",
    booktitle = {Protostars and Planets VII},
         year = 2023,
       editor = {{Inutsuka}, S. and {Aikawa}, Y. and {Muto}, T. and {Tomida}, K. and {Tamura}, M.},
       series = {Astronomical Society of the Pacific Conference Series},
       volume = {534},
        month = jul,
        pages = {1075},
       adsurl = {https://ui.adsabs.harvard.edu/abs/2023ASPC..534.1075N}
}

@INPROCEEDINGS{ceccarelli14,
       author = {{Ceccarelli}, C. and {Caselli}, P. and {Bockel{\'e}e-Morvan}, D. and {Mousis}, O. and {Pizzarello}, S. and {Robert}, F. and {Semenov}, D.},
        title = "{Deuterium Fractionation: The Ariadne's Thread from the Precollapse Phase to Meteorites and Comets Today}",
    booktitle = {Protostars and Planets VI},
         year = 2014,
       editor = {{Beuther}, Henrik and {Klessen}, Ralf S. and {Dullemond}, Cornelis P. and {Henning}, Thomas},
        month = jan,
        pages = {859},
          doi = {10.2458/azu_uapress_9780816531240-ch037},
archivePrefix = {arXiv},
       eprint = {1403.7143},
 primaryClass = {astro-ph.EP},
       adsurl = {https://ui.adsabs.harvard.edu/abs/2014prpl.conf..859C}
}

@ARTICLE{garrod11,
       author = {{Garrod}, R.~T. and {Pauly}, T.},
        title = "{On the Formation of CO$_{2}$ and Other Interstellar Ices}",
      journal = {\apj},
         year = 2011,
        month = jul,
       volume = {735},
       number = {1},
          eid = {15},
        pages = {15},
          doi = {10.1088/0004-637X/735/1/15},
archivePrefix = {arXiv},
       eprint = {1106.0540},
 primaryClass = {astro-ph.GA},
       adsurl = {https://ui.adsabs.harvard.edu/abs/2011ApJ...735...15G}
}

@ARTICLE{aikawa15,
       author = {{Aikawa}, Yuri and {Furuya}, Kenji and {Nomura}, Hideko and {Qi}, Chunhua},
        title = "{Analytical Formulae of Molecular Ion Abundances and the N$_{2}$H$^{+}$ Ring in Protoplanetary Disks}",
      journal = {\apj},
         year = 2015,
        month = jul,
       volume = {807},
       number = {2},
          eid = {120},
        pages = {120},
          doi = {10.1088/0004-637X/807/2/120},
archivePrefix = {arXiv},
       eprint = {1505.07550},
 primaryClass = {astro-ph.SR},
       adsurl = {https://ui.adsabs.harvard.edu/abs/2015ApJ...807..120A}
}

@ARTICLE{lupi21,
       author = {{Lupi}, Alessandro and {Bovino}, Stefano and {Grassi}, Tommaso},
        title = "{On the low ortho-to-para H$_{2}$ ratio in star-forming filaments}",
      journal = {\aap},
         year = 2021,
        month = oct,
       volume = {654},
          eid = {L6},
        pages = {L6},
          doi = {10.1051/0004-6361/202142145},
archivePrefix = {arXiv},
       eprint = {2109.02655},
 primaryClass = {astro-ph.GA},
       adsurl = {https://ui.adsabs.harvard.edu/abs/2021A&A...654L...6L}
}

@ARTICLE{bovino21,
       author = {{Bovino}, Stefano and {Lupi}, Alessandro and {Giannetti}, Andrea and {Sabatini}, Giovanni and {Schleicher}, Dominik R.~G. and {Wyrowski}, Friedrich and {Menten}, Karl M.},
        title = "{Chemical analysis of prestellar cores in Ophiuchus yields short timescales and rapid collapse}",
      journal = {\aap},
         year = 2021,
        month = oct,
       volume = {654},
          eid = {A34},
        pages = {A34},
          doi = {10.1051/0004-6361/202141252},
archivePrefix = {arXiv},
       eprint = {2105.02253},
 primaryClass = {astro-ph.GA},
       adsurl = {https://ui.adsabs.harvard.edu/abs/2021A&A...654A..34B}
}

@ARTICLE{LePetit16,
       author = {{Le Petit}, Franck and {Ruaud}, Maxime and {Bron}, Emeric and {Godard}, Benjamin and {Roueff}, Evelyne and {Languignon}, David and {Le Bourlot}, Jacques},
        title = "{Physical conditions in the central molecular zone inferred by H$_{3}$$^{+}$}",
      journal = {\aap},
         year = 2016,
        month = jan,
       volume = {585},
          eid = {A105},
        pages = {A105},
          doi = {10.1051/0004-6361/201526658},
archivePrefix = {arXiv},
       eprint = {1510.02221},
 primaryClass = {astro-ph.GA},
       adsurl = {https://ui.adsabs.harvard.edu/abs/2016A&A...585A.105L}
}

@ARTICLE{fuente23,
       author = {{Fuente}, A. and {Rivi{\`e}re-Marichalar}, P. and {Beitia-Antero}, L. and {Caselli}, P. and {Wakelam}, V. and {Esplugues}, G. and {Rodr{\'\i}guez-Baras}, M. and {Navarro-Almaida}, D. and {Gerin}, M. and {Kramer}, C. and {Bachiller}, R. and {Goicoechea}, J.~R. and {Jim{\'e}nez-Serra}, I. and {Loison}, J.~C. and {Ivlev}, A. and {Mart{\'\i}n-Dom{\'e}nech}, R. and {Spezzano}, S. and {Roncero}, O. and {Mu{\~n}oz-Caro}, G. and {Cazaux}, S. and {Marcelino}, N.},
        title = "{Gas phase Elemental abundances in Molecular cloudS (GEMS). VII. Sulfur elemental abundance}",
      journal = {\aap},
         year = 2023,
        month = feb,
       volume = {670},
          eid = {A114},
        pages = {A114},
          doi = {10.1051/0004-6361/202244843},
archivePrefix = {arXiv},
       eprint = {2212.03742},
 primaryClass = {astro-ph.GA},
       adsurl = {https://ui.adsabs.harvard.edu/abs/2023A&A...670A.114F}
}

@ARTICLE{crapsi05,
       author = {{Crapsi}, A. and {Caselli}, P. and {Walmsley}, C.~M. and {Myers}, P.~C. and {Tafalla}, M. and {Lee}, C.~W. and {Bourke}, T.~L.},
        title = "{Probing the Evolutionary Status of Starless Cores through N$_{2}$H$^{+}$ and N$_{2}$D$^{+}$ Observations}",
      journal = {\apj},
         year = 2005,
        month = jan,
       volume = {619},
       number = {1},
        pages = {379-406},
          doi = {10.1086/426472},
archivePrefix = {arXiv},
       eprint = {astro-ph/0409529},
 primaryClass = {astro-ph},
       adsurl = {https://ui.adsabs.harvard.edu/abs/2005ApJ...619..379C}
}

@ARTICLE{Chacon-Tanarro19,
       author = {{Chac{\'o}n-Tanarro}, A. and {Caselli}, P. and {Bizzocchi}, L. and {Pineda}, J.~E. and {Sipil{\"a}}, O. and {Vasyunin}, A. and {Spezzano}, S. and {Punanova}, A. and {Giuliano}, B.~M. and {Lattanzi}, V.},
        title = "{Mapping deuterated methanol toward L1544. I. Deuterium fraction and comparison with modeling}",
      journal = {\aap},
         year = 2019,
        month = feb,
       volume = {622},
          eid = {A141},
        pages = {A141},
          doi = {10.1051/0004-6361/201832703},
archivePrefix = {arXiv},
       eprint = {1808.09871},
 primaryClass = {astro-ph.GA},
       adsurl = {https://ui.adsabs.harvard.edu/abs/2019A&A...622A.141C}
}

@ARTICLE{keto10,
       author = {{Keto}, Eric and {Caselli}, Paola},
        title = "{Dynamics and depletion in thermally supercritical starless cores}",
      journal = {\mnras},
         year = 2010,
        month = mar,
       volume = {402},
       number = {3},
        pages = {1625-1634},
          doi = {10.1111/j.1365-2966.2009.16033.x},
archivePrefix = {arXiv},
       eprint = {0908.2400},
 primaryClass = {astro-ph.GA},
       adsurl = {https://ui.adsabs.harvard.edu/abs/2010MNRAS.402.1625K}
}

@ARTICLE{aikawa18,
       author = {{Aikawa}, Yuri and {Furuya}, Kenji and {Hincelin}, Ugo and {Herbst}, Eric},
        title = "{Multiple Paths of Deuterium Fractionation in Protoplanetary Disks}",
      journal = {\apj},
         year = 2018,
        month = mar,
       volume = {855},
       number = {2},
          eid = {119},
        pages = {119},
          doi = {10.3847/1538-4357/aaad6c},
archivePrefix = {arXiv},
       eprint = {1803.02498},
 primaryClass = {astro-ph.SR},
       adsurl = {https://ui.adsabs.harvard.edu/abs/2018ApJ...855..119A}
}

@ARTICLE{aikawa99,
       author = {{Aikawa}, Yuri and {Herbst}, Eric},
        title = "{Deuterium Fractionation in Protoplanetary Disks}",
      journal = {\apj},
         year = 1999,
        month = nov,
       volume = {526},
       number = {1},
        pages = {314-326},
          doi = {10.1086/307973},
       adsurl = {https://ui.adsabs.harvard.edu/abs/1999ApJ...526..314A}
}

@ARTICLE{hasegawa93,
       author = {{Hasegawa}, T.~I. and {Herbst}, E.},
        title = "{Three-Phase Chemical Models of Dense Interstellar Clouds - Gas Dust Particle Mantles and Dust Particle Surfaces}",
      journal = {\mnras},
         year = 1993,
        month = aug,
       volume = {263},
        pages = {589},
          doi = {10.1093/mnras/263.3.589},
       adsurl = {https://ui.adsabs.harvard.edu/abs/1993MNRAS.263..589H}
}

@Article{matplotlib,
  Author    = {Hunter, J. D.},
  Title     = {Matplotlib: A 2D graphics environment},
  Journal   = {Computing in Science \& Engineering},
  Volume    = {9},
  Number    = {3},
  Pages     = {90--95},
  publisher = {IEEE COMPUTER SOC},
  doi       = {10.1109/MCSE.2007.55},
  year      = 2007
}

@ARTICLE{furuya22,
       author = {{Furuya}, Kenji and {Hama}, Tetsuya and {Oba}, Yasuhiro and {Kouchi}, Akira and {Watanabe}, Naoki and {Aikawa}, Yuri},
        title = "{Diffusion Activation Energy and Desorption Activation Energy for Astrochemically Relevant Species on Water Ice Show No Clear Relation}",
      journal = {\apjl},
         year = 2022,
        month = jul,
       volume = {933},
       number = {1},
          eid = {L16},
        pages = {L16},
          doi = {10.3847/2041-8213/ac78e9},
archivePrefix = {arXiv},
       eprint = {2206.07225},
 primaryClass = {astro-ph.GA},
       adsurl = {https://ui.adsabs.harvard.edu/abs/2022ApJ...933L..16F}
}

@ARTICLE{taquet14,
       author = {{Taquet}, Vianney and {Charnley}, Steven B. and {Sipil{\"a}}, Olli},
        title = "{Multilayer Formation and Evaporation of Deuterated Ices in Prestellar and Protostellar Cores}",
      journal = {\apj},
         year = 2014,
        month = aug,
       volume = {791},
       number = {1},
          eid = {1},
        pages = {1},
          doi = {10.1088/0004-637X/791/1/1},
archivePrefix = {arXiv},
       eprint = {1405.3268},
 primaryClass = {astro-ph.GA},
       adsurl = {https://ui.adsabs.harvard.edu/abs/2014ApJ...791....1T}
}

@ARTICLE{pagani09,
       author = {{Pagani}, L. and {Vastel}, C. and {Hugo}, E. and {Kokoouline}, V. and {Greene}, C.~H. and {Bacmann}, A. and {Bayet}, E. and {Ceccarelli}, C. and {Peng}, R. and {Schlemmer}, S.},
        title = "{Chemical modeling of <ASTROBJ>L183</ASTROBJ> (<ASTROBJ>L134N</ASTROBJ>): an estimate of the ortho/para H\{\_2\} ratio}",
      journal = {\aap},
         year = 2009,
        month = feb,
       volume = {494},
       number = {2},
        pages = {623-636},
          doi = {10.1051/0004-6361:200810587},
archivePrefix = {arXiv},
       eprint = {0810.1861},
 primaryClass = {astro-ph},
       adsurl = {https://ui.adsabs.harvard.edu/abs/2009A&A...494..623P}
}

@ARTICLE{wakelam04,
       author = {{Wakelam}, V. and {Castets}, A. and {Ceccarelli}, C. and {Lefloch}, B. and {Caux}, E. and {Pagani}, L.},
        title = "{Sulphur-bearing species in the star forming region L1689N}",
      journal = {\aap},
         year = 2004,
        month = jan,
       volume = {413},
        pages = {609-622},
          doi = {10.1051/0004-6361:20031572},
archivePrefix = {arXiv},
       eprint = {astro-ph/0310328},
 primaryClass = {astro-ph},
       adsurl = {https://ui.adsabs.harvard.edu/abs/2004A&A...413..609W}
}

@ARTICLE{hocuk17,
       author = {{Hocuk}, S. and {Sz{\H{u}}cs}, L. and {Caselli}, P. and {Cazaux}, S. and {Spaans}, M. and {Esplugues}, G.~B.},
        title = "{Parameterizing the interstellar dust temperature}",
      journal = {\aap},
         year = 2017,
        month = aug,
       volume = {604},
          eid = {A58},
        pages = {A58},
          doi = {10.1051/0004-6361/201629944},
archivePrefix = {arXiv},
       eprint = {1704.02763},
 primaryClass = {astro-ph.GA},
       adsurl = {https://ui.adsabs.harvard.edu/abs/2017A&A...604A..58H}
}

@ARTICLE{cazaux17,
       author = {{Cazaux}, S. and {Mart{\'\i}n-Dom{\'e}nech}, R. and {Chen}, Y.~J. and {Mu{\~n}oz Caro}, G.~M. and {Gonz{\'a}lez D{\'\i}az}, C.},
        title = "{CO Depletion: A Microscopic Perspective}",
      journal = {\apj},
         year = 2017,
        month = nov,
       volume = {849},
       number = {2},
          eid = {80},
        pages = {80},
          doi = {10.3847/1538-4357/aa8b0c},
archivePrefix = {arXiv},
       eprint = {1709.01285},
 primaryClass = {astro-ph.SR},
       adsurl = {https://ui.adsabs.harvard.edu/abs/2017ApJ...849...80C}
}

@ARTICLE{molpeceres20,
       author = {{Molpeceres}, Germ{\'a}n and {K{\"a}stner}, Johannes},
        title = "{Adsorption of H2 on amorphous solid water studied with molecular dynamics simulations}",
      journal = {Physical Chemistry Chemical Physics (Incorporating Faraday Transactions)},
         year = 2020,
        month = apr,
       volume = {22},
       number = {14},
        pages = {7552-7563},
          doi = {10.1039/D0CP00250J},
archivePrefix = {arXiv},
       eprint = {2003.08873},
 primaryClass = {physics.chem-ph},
       adsurl = {https://ui.adsabs.harvard.edu/abs/2020PCCP...22.7552M}
}

@ARTICLE{furuya15,
       author = {{Furuya}, K. and {Aikawa}, Y. and {Hincelin}, U. and {Hassel}, G.~E. and {Bergin}, E.~A. and {Vasyunin}, A.~I. and {Herbst}, E.},
        title = "{Water deuteration and ortho-to-para nuclear spin ratio of H$_{2}$ in molecular clouds formed via the accumulation of H I gas}",
      journal = {\aap},
         year = 2015,
        month = dec,
       volume = {584},
          eid = {A124},
        pages = {A124},
          doi = {10.1051/0004-6361/201527050},
archivePrefix = {arXiv},
       eprint = {1510.05135},
 primaryClass = {astro-ph.GA},
       adsurl = {https://ui.adsabs.harvard.edu/abs/2015A&A...584A.124F}
}

@ARTICLE{nagaoka05,
       author = {{Nagaoka}, Akihiro and {Watanabe}, Naoki and {Kouchi}, Akira},
        title = "{H-D Substitution in Interstellar Solid Methanol: A Key Route for D Enrichment}",
      journal = {\apjl},
         year = 2005,
        month = may,
       volume = {624},
       number = {1},
        pages = {L29-L32},
          doi = {10.1086/430304},
archivePrefix = {arXiv},
       eprint = {astro-ph/0503587},
 primaryClass = {astro-ph},
       adsurl = {https://ui.adsabs.harvard.edu/abs/2005ApJ...624L..29N}
}

@ARTICLE{hidaka09,
       author = {{Hidaka}, H. and {Watanabe}, M. and {Kouchi}, A. and {Watanabe}, N.},
        title = "{Reaction Routes in the CO-H$_{2}$CO-d$_{n}$ -CH$_{3}$OH-d$_{m}$ System Clarified from H(D) Exposure of Solid Formaldehyde at Low Temperatures}",
      journal = {\apj},
         year = 2009,
        month = sep,
       volume = {702},
       number = {1},
        pages = {291-300},
          doi = {10.1088/0004-637X/702/1/291},
       adsurl = {https://ui.adsabs.harvard.edu/abs/2009ApJ...702..291H}
}

@ARTICLE{furuya19,
       author = {{Furuya}, Kenji and {Aikawa}, Yuri and {Hama}, Tetsuya and {Watanabe}, Naoki},
        title = "{H$_{2}$ Ortho-Para Spin Conversion on Inhomogeneous Grain Surfaces}",
      journal = {\apj},
         year = 2019,
        month = sep,
       volume = {882},
       number = {2},
          eid = {172},
        pages = {172},
          doi = {10.3847/1538-4357/ab3790},
archivePrefix = {arXiv},
       eprint = {1908.01966},
 primaryClass = {astro-ph.GA},
       adsurl = {https://ui.adsabs.harvard.edu/abs/2019ApJ...882..172F}
}

@ARTICLE{he16,
       author = {{He}, Jiao and {Acharyya}, Kinsuk and {Vidali}, Gianfranco},
        title = "{Sticking of Molecules on Nonporous Amorphous Water Ice}",
      journal = {\apj},
         year = 2016,
        month = may,
       volume = {823},
       number = {1},
          eid = {56},
        pages = {56},
          doi = {10.3847/0004-637X/823/1/56},
archivePrefix = {arXiv},
       eprint = {1602.06341},
 primaryClass = {astro-ph.IM},
       adsurl = {https://ui.adsabs.harvard.edu/abs/2016ApJ...823...56H}
}

@ARTICLE{amiaud15,
       author = {{Amiaud}, Lionel and {Fillion}, Jean-Hugues and {Dulieu}, Fran{\c{c}}ois and {Momeni}, Anouchah and {Lemaire}, Jean-Louis},
        title = "{Physisorption and desorption of H2, HD and D2on amorphous solid water ice. Effect on mixing isotopologue on statistical population of adsorption sites}",
      journal = {Physical Chemistry Chemical Physics (Incorporating Faraday Transactions)},
         year = 2015,
        month = jan,
       volume = {17},
       number = {44},
        pages = {30148-30157},
          doi = {10.1039/C5CP03985A},
       adsurl = {https://ui.adsabs.harvard.edu/abs/2015PCCP...1730148A}
}
\bibliographystyle{aasjournal}



\end{document}